\documentclass[%
pre,
aps,
a4paper,
english,
reprint,
twocolumn,
superscriptaddress,showkeys
]{revtex4-2}

\usepackage[T1]{fontenc}
\usepackage[utf8]{inputenc}
\usepackage{amsmath}
\usepackage{amssymb}
\usepackage{enumerate}
\usepackage{graphicx}
\usepackage{bm}
\usepackage{refstyle}
\usepackage{multirow,array}
\usepackage{mathptmx}
\usepackage{mathtools}
\usepackage{empheq}
\usepackage{chessfss}
\usepackage{epsdice}
\usepackage{pifont}
\usepackage{amsthm}
\usepackage{bbold}
\usepackage{cancel}
\usepackage{makecell}
\usepackage[
citecolor=blue,
colorlinks,
linkcolor=blue,
urlcolor=blue,
]{hyperref}
\usepackage[dvipsnames]{xcolor}
\usepackage{dcolumn}
\usepackage{bm}
\newcommand{\indsize}{\scriptsize}
\newcommand{\colind}[2]{\displaystyle\smash{\mathop{#1}^{\raisebox{.5\normalbaselineskip}{\indsize #2}}}}
\newcommand{\rowind}[1]{\mbox{\indsize #1}}

\begin{document}

\title{Upstream reciprocity versus downstream reciprocity: Catalyzing cooperation}

\author{Vikash Kumar Dubey}
\email{vdubey9818@gmail.com}
\affiliation{
	Department of Physics,
	Indian Institute of Technology Kanpur,
	Uttar Pradesh 208016, India
}
	\author{Sagar Chakraborty}
	\email{sagarc@iitk.ac.in}
	\affiliation{
		Department of Physics,
		Indian Institute of Technology Kanpur,
		Uttar Pradesh 208016, India
	}
	\begin{abstract}
Why would anyone help a stranger, knowing they may never meet again? Indirect reciprocity offers one of the most compelling evolutionary answers, yet its two canonical forms---upstream reciprocity (experience-based), and downstream reciprocity (reputation-based)---have been studied mostly in isolation. Their joint dynamics in finite and structured populations remain largely unexplored.
Here, we fill this gap using agent-based simulations in which an agent is behaviourally either defector, upstream reciprocator, or downstream reciprocator, and the agents' population state is temporally updated using different evolutionary update mechanisms. We show that update mechanism plays a surprisingly decisive role in shaping the fate of downstream and especially upstream reciprocators. Whether agents' experiences and reputations are updated globally or locally can shift outcomes from rich behavioural coexistence to the dominance of downstream reciprocators alone. Intriguingly, we uncover a robust structural feature that persists across all the explored update rules and population sizes: an optimal network degree at which upstream reciprocity is maximized, reflecting a fundamental tug-of-war between cooperative clustering and exposure to defectors. Our results highlight that while downstream reciprocity can either foster or inhibit upstream reciprocity depending on the update mechanism, its net effect on cooperation remains largely positive.
	\end{abstract}
	\keywords{Indirect reciprocity, Cooperation, Replicator dynamics, Evolutionary games on networks, Finite populations}
	\maketitle
	
\section{Introduction}
	
The evolution of cooperation among unrelated individuals has long been a central puzzle in evolutionary theory~\cite{CluttonBrock2009}. In many real-world scenarios, individuals help strangers whom they are meeting for the first time and may never meet again. Such behaviors are often explained by indirect reciprocity, one of the central mechanisms catalyzing cooperation among strangers~\cite{Nowak2006_five_rules}. Empirical studies have documented indirect reciprocity not only in humans~\cite{Dufwenberg2001,Bartlett2006,Gray2014,vanApeldoorn2016,Horita2016,Schnedler2022,Obayashi2023,Watanabe2014,Swakman2016,Eriksson2021,Gaudeul2021} but also in several animal species~\cite{Rutte2007,Bshary2006,Majolo2012,Russell2008}, suggesting that cooperation driven by indirect reciprocity~\cite{Nowak2005} is not an exclusive signature of cognitively advanced species like humans, but a far more widespread feature of the biological world.

Indirect reciprocity mainly works on two distinct mechanisms: upstream reciprocity~\cite{Boyd1988,Boyd1989,Nowak2006,Iwagami2010} and downstream reciprocity~\cite{Nowak1998a,Nowak1998b,Ohtsuki2004,Nowak2005,Hilbe2018,Santos2018}. Under upstream reciprocity, an individual who has received help is driven by a sense of gratitude (or a `pay-it-forward' mentality) to subsequently help a third party. A behavior often summarized by the phrase: "You scratch my back and I’ll scratch someone else’s". Under downstream reciprocity, individuals preferentially help those with a good reputation, anticipating that such behavior would enhance their own likelihood of receiving help in the future. This mechanism is often summarized by the aphorism: “I scratch your back and someone else will scratch mine.”

A vast body of literature has investigated the role of downstream reciprocity in promoting cooperation. Early studies emphasized the importance of reputation in shaping social interactions~\cite{Raub1990,Pollock1992}. However, the field developed rapidly following the celebrated work of Nowak and Sigmund~\cite{Nowak1998a,Nowak1998b}, who introduced a formal mathematical framework for downstream reciprocity based on image-scoring and identified conditions under which it can sustain cooperation. Subsequently, Ohtsuki and Iwasa~\cite{Ohtsuki2004,Ohtsuki2006_IR} extended this framework by analyzing higher-order social norms, showing that cooperative behavior toward a recipient can be considered good or bad depending on the reputation of the recipient and the socially accepted norm. Their analysis revealed a small set of so-called "leading eight" norms that are evolutionarily stable and capable of sustaining indirect reciprocity under a wide range of conditions~\cite{Hilbe2018,Nowak2005,Santos2018,Schmid2021}.

In contrast, upstream reciprocity is comparatively less explored. Early studies showed that upstream reciprocity can emerge in small groups~\cite{Boyd1988,Pfeiffer2005}. Subsequent work demonstrated that factors such as relatedness~\cite{Nowak2006}, network structure and direct reciprocity can facilitate the emergence of cooperation through upstream reciprocity~\cite{Nowak2006,Iwagami2010}.

It is well known that downstream reciprocity can outperform defectors, as reputation systems allow individuals to identify those who defect. By tagging defectors as “bad,” downstream reciprocators (individuals employing reputation-based downstream reciprocity exclusively) reduce defectors' access to cooperative benefits, thereby limiting their evolutionary success~\cite{Nowak2005,Okada2020}. In contrast, the evolutionary viability of upstream reciprocity against defectors is less straightforward~\cite{Okada2020,Sigmund2010}. Because upstream reciprocators (individuals employing experience-based upstream reciprocity exclusively) condition their behavior solely on personal experience rather than public reputation, defectors are not directly excluded from receiving a fitness advantage, leaving the population highly vulnerable to exploitation by the latter. 

These two mechanisms have been mostly studied in isolation, and only a limited number of studies have considered them jointly. Some experimental studies have analyzed both upstream and downstream reciprocity within a single experimental framework~\cite{Baker2014,Stanca2009,Simpson2017}, providing evidence that one mechanism can reinforce the other~\cite{Simpson2017}. 
Motivated by these findings, recent theoretical work has begun to investigate their joint dynamics within a unified framework. Notably, Ref.~\cite{Sasaki2024} developed a theoretical model that integrates both upstream and downstream behaviors within a single individual and, using replicator dynamics, showed that integrated reciprocators---combining both mechanisms---can coexist with unconditional cooperators and defectors. This framework was later extended to multiplayer public goods games~\cite{Sasaki2026}, where such coexistence was again demonstrated.

Real populations are, however, finite and structured. It is therefore natural to ask how upstream and downstream reciprocity evolve when analyzed jointly in finite, structured populations, where individuals interact with only a limited number of others. Moreover, in such populations, the rules governing how individuals update their strategies, reputations, and personal experiences can vary widely---specifically, whether reputations and personal experiences may get updated either globally (across all individuals simultaneously) or locally (only among a subset of individuals). Given our results in the paper show that this distinction can fundamentally change evolutionary outcomes.
It is surprising  that this dimension has received little attention in the literature.

Aforementioned setup allows us to address several key questions: Can downstream reciprocity promote upstream reciprocity in a population subject to invasion by defectors? Do the two mechanisms reinforce each other, as suggested by experimental findings~\cite{Simpson2017}, or do they suppress one another? Does the choice of update rule---global versus local---fundamentally alter these outcomes? And what is the role of network structure and size in governing the evolutionary fate of these behaviors? Despite the widespread relevance of these questions, they remain poorly understood. In this paper, we aim to address this lacuna by studying the joint evolution of upstream and downstream reciprocity in finite structured populations under multiple update rules. Doing so, we argue, is a necessary step toward completing the picture of how indirect reciprocity evolves and endures in human and animal societies.

The paper is organized as follows: We begin in Sec.~\ref{sec:INF_model} with an infinite-population model which, while unable to capture the full complexity of finite structured populations, provides a useful analytical benchmark and coarse-grained intuition for the outcomes to expect. The finite-population model---which forms the core of this work---is then constructed in Sec.~\ref{sec:concept_framework}, where we describe the system conceptually. Subsequently, its key assumptions and numerical architecture are discussed in details in Sec.~\ref{sec:numercial_scheme}. Using this framework, we analyze the evolutionary dynamics under different update rules in Sec.~\ref{sec:Results}, demonstrating how global and local updating mechanisms play an important role in governing evolutionary outcomes. We synthesize and compare the results across update rules in Sec.~\ref{sec:comphrendhing}, before concluding and outlining future directions in Sec.~\ref{sec:conclusion}.

\section{Infinite Population Limit: An Analytical Prelude}\label{sec:INF_model}

In this section, we construct a simple deterministic, compartmental model~\cite{Chang2020} to gain analytical insight into how downstream reciprocity facilitates the emergence of upstream reciprocators in the presence of defectors. In particular, using replicator-mutator dynamics~\cite{hofbauer_book,Nowak_2006_book,Mukhopadhyay2021}, we demonstrate that, depending on the parameter values, the system can exhibit the following dynamical outcomes: (i) a stable coexistence among upstream reciprocators, downstream reciprocators, and defectors; (ii) a stable monomorphic state of downstream reciprocators alone; or (iii) bistability between this coexistence and a downstream-dominated state. In contrast, in the absence of downstream individuals, upstream reciprocity alone fails to catalyze cooperation, and the system is invariably dominated by defectors. This preliminary analysis sets the stage for our subsequent investigation in finite and structured populations.

At this point, we note that Ref.~\cite{Sasaki2024} evaluates a similar infinite-population framework and likewise predicts the coexistence of a reciprocating strategy with cooperators and defectors. While Ref.~\cite{Sasaki2024} primarily focuses on the behavior of an integrated reciprocator (which merge both downstream and upstream behavior together), our objective is to isolate reputation-based and experience-based decision making into distinct strategies, allowing us to systematically examine how these two phenomena interact and affect each other. Furthermore, the presence of unconditional cooperators in their framework may artificially sustain reciprocators; we choose to examine whether reciprocators remain viable in the harsher environment of upstream reciprocators, downstream reciprocators, and defectors alone, with no unconditional cooperators present.

Keeping this in mind, let us consider an infinite, well-mixed population consisting of three behavioural types: upstream reciprocators, downstream reciprocators, and defectors. We denote their respective frequencies as $x_1$, $x_2$, and $x_3$, such that $x_1 + x_2 + x_3 = 1$. Upstream individuals condition their actions entirely on personal experience, dictated by an internal level of emotional gratitude: a fraction $h$ possess high gratitude and cooperate, while the remaining fraction $1-h$ possess low gratitude and defect. Conversely, downstream reciprocators rely on public reputation, cooperating only when interacting with an individual of good image and defecting otherwise. Finally, defectors unconditionally defect across all encounters.

To formalize these image-based interactions, let $g_{\mathrm U}$ and $g_{\mathrm{Dn}}$ denote the fractions of upstream and downstream individuals possessing a good image, respectively; consequently, the remaining fractions $1 - g_{\mathrm U}$ and $1 - g_{\mathrm{Dn}}$ represent those with a bad image. Note that while one could generally assign a good-image fraction $g_{\mathrm D}$ to the defector subpopulation, their persistent non-cooperative behavior rapidly degrades their public reputation. Therefore, in the present model, we naturally fix $g_{\mathrm D} = 0$, reflecting a scenario where defectors are universally recognized as having a bad image and are thus completely barred from receiving downstream cooperation.

When two individuals interact, a cooperating individual incurs a cost $c$ and provides a benefit $b$ to the opponent, whereas defection yields a payoff of zero to both individuals. Let us now compute the payoff matrix in terms of these parameters. Consider an interaction between two upstream reciprocators. A randomly chosen upstream reciprocator possesses high emotional gratitude with probability $h$ (and thus cooperates) and low emotional gratitude with probability $1-h$ (and thus defects). The same probabilities apply independently to their opponent. Consequently, the interaction falls into one of four scenarios depending on the types of both players:
\begin{enumerate}[(i)]
	\item With probability $h^2$, both individuals possess high emotional gratitude and both will cooperate, meaning both receive the benefit $b$ and pay the cost $c$. Therefore, the payoff received by the focal player is $b-c$.
	\item With probability $h(1-h)$, the focal individual has high gratitude while the opponent has low gratitude; therefore focal individual will cooperate and the opponent will defect. This means that the focal player pays the cost $c$ and receives no benefit.
	\item With probability $(1-h)h$, the focal individual has low gratitude while the opponent has high gratitude; therefore focal individuals will defect and the opponent will cooperate. This implies that the focal individual receives the benefit $b$ and pays no cost.
	\item With probability $(1-h)^2$, both possess low gratitude; therefore both will defect and receive zero benefit.
\end{enumerate}
Summing over these possibilities, the expected payoff $\pi(\rm{U},\rm{U})$ of an upstream individual interacting with another upstream individual is given by the following weighted average:
\begin{eqnarray}
	\pi(\rm{U},\rm{U}) &=& h^2(b-c) - h(1-h)c + (1-h)hb + (1-h)^2 \cdot 0 \nonumber \\
	&=& h(b-c).
\end{eqnarray}

Similarly, consider the interaction between an upstream and a downstream reciprocators.
A randomly chosen upstream reciprocator has a good image with probability $g_{\mathrm U}$ and a bad image with probability $1-g_{\mathrm U}$. Assuming independence between emotional state and image, the joint probabilities are given as follows: 
\begin{enumerate}[(i)]
\item With probability $hg_{\mathrm U}$, the upstream reciprocator has high emotional gratitude and a good image; therefore, the downstream reciprocator will cooperate and pay a cost $c$ because they are meeting an opponent with a good image. The upstream reciprocator will also cooperate because of their high gratitude. This gives the focal upstream reciprocator a payoff of $b-c$. 

\item With probability $h(1-g_{\mathrm U})$, the upstream reciprocator has high emotional gratitude and a bad image; therefore, the upstream reciprocator will cooperate and pay the cost $c$, whereas the downstream reciprocator provides no benefit because it defects.

\item With probability $(1-h)g_{\mathrm U}$, the upstream reciprocator has low emotional gratitude and a good image; therefore, the upstream reciprocator defects and pays no cost, whereas the downstream reciprocator provides the benefit $b$ to the focal upstream reciprocator because of its good image. 

\item With probability $(1-h)(1-g_{\mathrm U})$, the upstream reciprocator has low gratitude and a bad image; therefore, both individuals will defect and receive a payoff of zero. 
\end{enumerate}
Consequently, the expected payoff $\pi(\rm{U},\rm{D})$ of an upstream individual interacting with a downstream individual is given by:
\begin{eqnarray}
	\pi(\rm{U},\rm{Dn}) &=& hg_{\mathrm U}(b-c) + (1-h)g_{\mathrm U} b - h(1-g_{\mathrm U})c \nonumber \\&&+ (1-h)(1-g_{\mathrm U})\cdot 0 \nonumber \\
	&=& bg_{\mathrm U} - h c.
\end{eqnarray}

Next, consider the interaction between an upstream reciprocator and a defector. When facing a defector, an upstream individual will only cooperate if they possess high emotional gratitude, incurring a cost $c$. Because a defector never cooperates, the upstream individual receives no benefit in return. Therefore, the expected payoff of upstream individual interacting defector is:
\begin{equation}
	\pi({\rm U},{\rm D}) = -hc.
\end{equation}

We can calculate the payoffs involving downstream reciprocators in a similar manner. When a downstream individual interacts with an upstream reciprocator, the downstream player cooperates only if the upstream partner has a good image (which occurs with probability $g_{\mathrm U}$). The expected payoff of downstream reciprocator is therefore:
\begin{eqnarray}
	\pi(\rm{Dn},\rm{U}) &=& hg_{\mathrm U}(b-c) + h(1-g_{\mathrm U}) b - (1-h)g_{\mathrm U}c \nonumber \\&&+ (1-h)(1-g_{\mathrm U})\cdot 0 \nonumber \\
	&=&hb-cg_{\mathrm U}.
\end{eqnarray}
Similarly, the expected payoff of a downstream reciprocator interacting with another downstream reciprocator is given by:
\begin{eqnarray}
	\pi(\rm{Dn},\rm{Dn})&=& g_{\mathrm{Dn}}^2(b-c) + g_{\mathrm{Dn}}(1-g_{\mathrm{Dn}}) b - (1-g_{\mathrm{Dn}})g_{\mathrm{Dn}}c \nonumber\\ &&+ (1-g_{\mathrm{Dn}})^2\cdot 0 \nonumber \\
	&=&(b-c)g_{\mathrm{Dn}}.
\end{eqnarray}
When a downstream reciprocator interacts with a defector, it receive a payoff of zero. Because downstream individuals condition their behavior on public reputation, they can identify defectors, as defectors have a bad image ($g_{\rm D}=0$), and refuse to cooperate, yielding $\pi(\rm{Dn},\rm{D})=0$.

Finally, since defectors never cooperate, neither costs nor benefits are realized in interactions among defectors themselves or with downstream reciprocators. However, a defector can obtain a positive payoff when interacting with an upstream individual possessing high emotional gratitude. Therefore, $\pi({\rm D},{\rm U})=bh$, while $\pi(\rm{D},\rm{Dn})=\pi(\rm{D},\rm{D})=0$.

The above discussion leads to the following payoff matrix
\begin{equation}
	{\sf \Pi}=\begin{array}{@{}c@{}}\label{eqn:payoff3by3}
		\rowind{U} \\ \rowind{Dn} \\ \rowind{D} \\ 
	\end{array}
	\mathop{\left[
		\begin{array}{ *{5}{c} }
			\colind{h (b-c)}{U}  &  \colind{b g_{\mathrm U}-c h}{Dn}  &  \colind{-ch}{D}   \\
			b h-c g_{\mathrm U}&   (b-c) g_{\mathrm{Dn}}  &0 \\
			b h&0 &0
		\end{array}
		\right]}^{
		\begin{array}{@{}c@{}}
			\\ \mathstrut
		\end{array}
	}.
\end{equation}

We utilize the standard replicator–mutator (also called selection–mutation) dynamics~\cite{hofbauer_book,Nowak_2006_book,Mukhopadhyay2021} to determine the fixed points of the system and analyze their stability:
\begin{equation}\label{eqn:replicator_mut}
	\dot{x}_i = x_i (f_i - \phi) + \mu \left( 1 - 3 x_i \right), \quad i\in\{1, 2, 3\},
\end{equation}
where indices $1$, $2$, and $3$ correspond to upstream reciprocators, downstream reciprocators, and defectors, respectively. Here, $f_i = \sum_j x_j {\sf \Pi}_{ij}$ denotes the expected fitness of the $i$-th behavioural type, $\phi = \sum_{i,j} x_i x_j {\sf \Pi}_{ij}$ represents the average fitness of the entire population, and $\mu$ scales the uniform, additive mutation rate across strategies.

To explicitly demonstrate the effect of downstream reciprocators, we first consider a reduced system in which only upstream reciprocators and defectors are present. The corresponding reduced payoff matrix is given by:
\begin{equation}\label{eqn:payoff2by2}
	{\sf \Pi}^{\rm r} =
	\begin{array}{c}
	\rowind{U} \\  \rowind{D} \\ 
	\end{array}
	\left[
	\begin{array}{cc}
		\colind{h(b-c)}{U} & \colind{-ch}{D} \\
		bh & 0
	\end{array}
	\right].
\end{equation}
For this two-strategy system, the replicator--mutator equation reduces to:
\begin{equation}
	\dot{x} = x(1-x)\left[f_{\rm{U}}(x) - f_{\rm D}(x)\right] + \mu(1 - 2x),
\end{equation}
where $x$ denotes the frequency of upstream reciprocators, leaving $1-x$ as the frequency of defectors. The expected fitnesses are given by $f_{\rm{U}}(x)\equiv x{\sf \Pi}^{\rm r}_{11}+(1-x){\sf \Pi}^{\rm r}_{12} = xbh - ch$ and $f_{\rm{D}}(x)\equiv x{\sf \Pi}^{\rm r}_{21}+(1-x){\sf \Pi}^{\rm r}_{22} = xbh$. Substituting these expressions yields the simplified dynamical equation:
\begin{equation}\label{eq:reduced_sys}
	\dot{x} = -ch(1-x)x + \mu(1 - 2x).
\end{equation}
The fixed point of this equation in the interval \( (0,1) \) is given by
\begin{equation}
	x^* = \frac{ch + 2\mu - \sqrt{c^2 h^2 + 4\mu^2}}{2ch}.
\end{equation}

It is straightforward to verify that for small mutation rates ($\mu \ll 1$), this fixed point lies close to $x=0$, indicating a state heavily dominated by defectors. To determine the stability of this equilibrium, we compute the Jacobian:
\begin{equation}
	J(x^*) = ch(2x^* - 1) - 2\mu = -\sqrt{c^2 h^2 + 4\mu^2}.
\end{equation}
Since $J(x^*) < 0$, the fixed point is locally asymptotically stable. We thus conclude that in the absence of downstream reputation mechanisms, upstream reciprocators cannot robustly emerge, and defectors inevitably dominate the population.

Let us now consider the full system in which all three behavioural types are present, with the payoff matrix given by Eq.~(\ref{eqn:payoff3by3}). To determine the fixed points of Eq.~(\ref{eqn:replicator_mut}), we must solve the resulting system of coupled nonlinear equations. Because analytical solutions are difficult to obtain in closed form, we employ a perturbative approach and compute the fixed points up to first order in $\mu$. 

We start by assuming that the perturbed fixed point takes the form:
\begin{equation}\label{eqn:x1x2_perturb}
	(\hat{x}_1,\hat{x}_2) = (x_1^* + \mu \xi_1, x_2^* + \mu \xi_2),
\end{equation}
where $(x_1^*, x_2^*)$ is a valid equilibrium of the system in the absence of mutation ($\mu = 0$). Note that we suppress the third coordinate using the relation $x_3 = 1 - x_1 - x_2$. The four valid roots for the unperturbed system ($\mu = 0$) are given by the trivial vertices $(0,0)$, $(0,1)$, $(1,0)$, and the internal coexisting fixed point is given by $ \left(\frac{(b-c)g_{\mathrm{Dn}} h}{b g_{\mathrm U}^2}, \frac{c h}{b g_{\mathrm U}}\right)$. To solve for the first-order correction terms $\xi_1$ and $\xi_2$, we can express Eq.~(\ref{eqn:replicator_mut}) as:
\begin{equation}
	\dot{x}_1 = F_1(x_1,x_2) + \mu F_2(x_1)~\text{and}~
	\dot{x}_2 = G_1(x_1,x_2) + \mu G_2(x_2),
\end{equation}
where $F_1 = x_1(f_1 - \phi)$, $F_2=1-3x_1$, $G_1 = x_2(f_2 - \phi)$ and $G_2=1-3x_2$. 
Our goal is to determine $(\hat{x}_1,\hat{x}_2)$ such that:
\begin{equation}
F_1(\hat{x}_1,\hat{x}_2) + \mu F_2(\hat{x}_1) = 0~\text{and}~ 
G_1(\hat{x}_1,\hat{x}_2) + \mu G_2(\hat{x}_2) = 0.
\end{equation}
By substituting the perturbation ansatz [Eq.~(\ref{eqn:x1x2_perturb})] into these expressions and performing a Taylor expansion up to first order in $\mu$, we obtain a system of linear equations for the corrections $\xi_1$ and $\xi_2$. Solving this system yields the following first-order perturbed solutions:
\begin{subequations}\label{eqn:perturb_sol}
	\begin{align}
		\hat{x}_1 &= x_1^* + \mu 
		\frac{-F_2(x_1^*) \frac{\partial G_1}{\partial x_2} 
			+ G_2(x_2^*) \frac{\partial F_1}{\partial x_2}}
		{\frac{\partial F_1}{\partial x_1}\frac{\partial G_1}{\partial x_2}
			- \frac{\partial G_1}{\partial x_1}\frac{\partial F_1}{\partial x_2}}, \\
		\hat{x}_2 &= x_2^* + \mu 
		\frac{-F_2(x_1^*) \frac{\partial G_1}{\partial x_1} 
			+ G_2(x_2^*) \frac{\partial F_1}{\partial x_1}}
		{\frac{\partial F_1}{\partial x_2}\frac{\partial G_1}{\partial x_1}
			- \frac{\partial G_1}{\partial x_2}\frac{\partial F_1}{\partial x_1}}.
	\end{align}
\end{subequations}
where all partial derivatives are evaluated at the unperturbed fixed point $(x_1^*, x_2^*)$.

\begin{figure}[h]
	\includegraphics[scale=0.43]{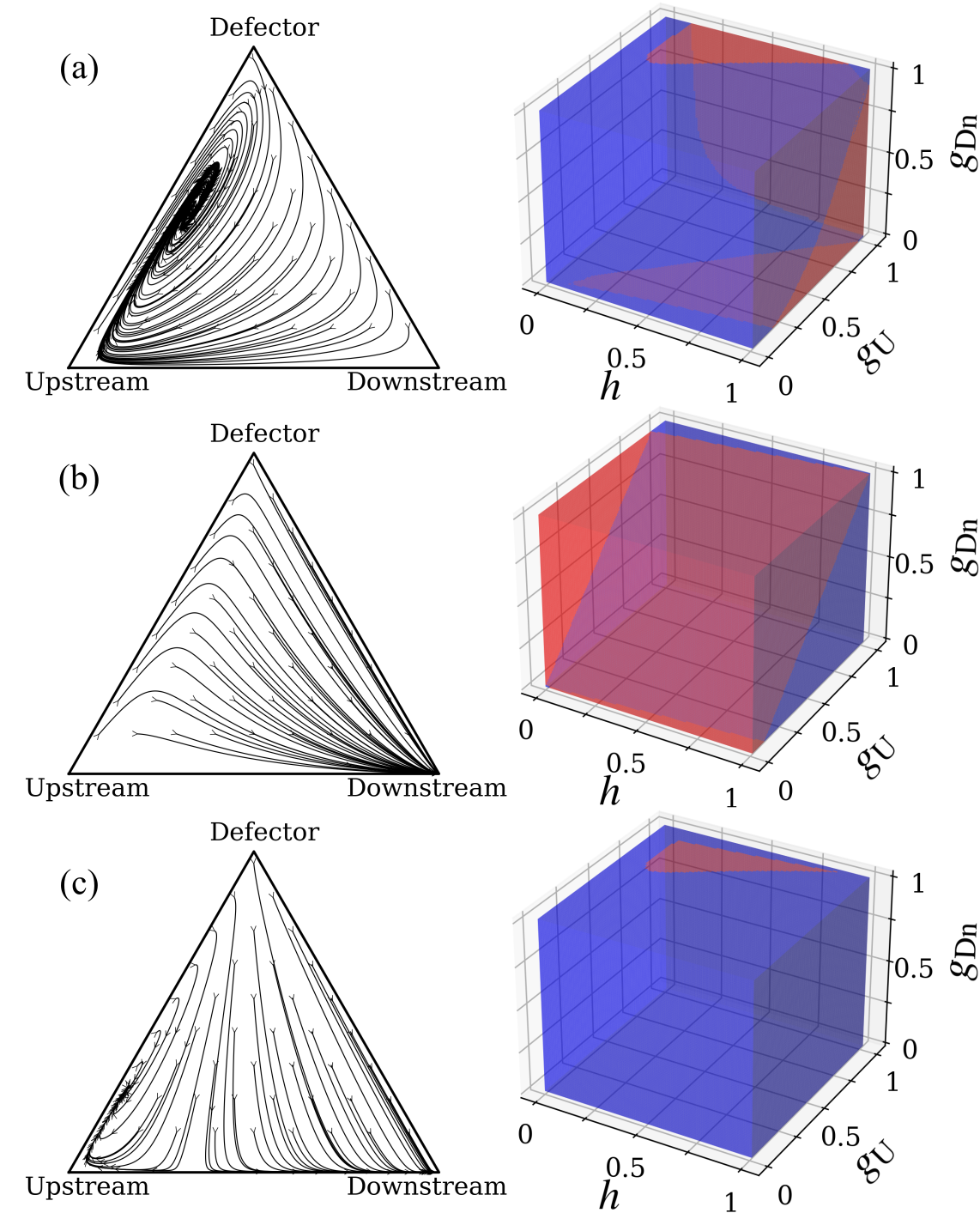}
	\caption{Regions of existence and stability of fixed points in the parameter space $(h, g_{\mathrm U}, g_{\mathrm{Dn}})$. The left panels (a–c) illustrate the dynamical trajectories on the strategy simplex for representative parameter choices, while the right panel displays the broader parameter regions where these distinct dynamics occur. The red volume denotes the parameter space where the fixed point(s) lies within the simplex and is stable. The blue region corresponds to parameter values where these conditions (the fixed point stability and lying within the simplex) are not satisfied. The specific parameter choices in the left panels are: (a) $h = 0.7$, $g_{\mathrm U} = 0.8$, $g_{\mathrm{Dn}} = 0.3$; (b) $h = 0.5$, $g_{\mathrm U} = 0.4$, $g_{\mathrm{Dn}} = 0.9$; and (c) $h = 0.2$, $g_{\mathrm U} = 0.8$, $g_{\mathrm{Dn}} = 1.0$.
	}
	\label{fig:replicator_mutator_model}
\end{figure}

Using Eq.~(\ref{eqn:perturb_sol}), the internal fixed point is can be evaluated as:
\begin{eqnarray}
	&&\hat{x}_1 =
	\frac{g_{\mathrm{Dn}} h (b-c)}{b g_{\mathrm U}^2} +\nonumber\\
	 &&\mu \frac{b^2 (g_{\mathrm{Dn}} h - g_{\mathrm U}^2) + b c (-3 g_{\mathrm{Dn}} h + g_{\mathrm U}^2 + 2 g_{\mathrm U} h) + c^2 h (2 g_{\mathrm{Dn}} - g_{\mathrm U})}
	{c^2 h \left[b (g_{\mathrm{Dn}} h - g_{\mathrm U}^2) + c h (g_{\mathrm U} - g_{\mathrm{Dn}})\right]}, \nonumber\\
&&\hat{x}_2 = \frac{c h}{b g_{\mathrm U}} + \mu \frac{g_{\mathrm U} \left[b (g_{\mathrm U}^2 - 2 g_{\mathrm{Dn}} h) + c h (2 g_{\mathrm{Dn}} - g_{\mathrm U})\right]}
	{g_{\mathrm{Dn}} h (b-c) \left[b (g_{\mathrm{Dn}} h - g_{\mathrm U}^2) + c h (g_{\mathrm U} - g_{\mathrm{Dn}})\right]}.
\end{eqnarray}
Similarly, the fixed point near the vertex $(0,1,0)$, corresponding to the downstream-dominated region, is given by:
\begin{eqnarray}
	\hat{x}_1 &=& \mu \frac{1}{b (g_{\mathrm{Dn}} - g_{\mathrm U}) + c (h - g_{\mathrm{Dn}})}, \nonumber\\
	\hat{x}_2 &=& 1 + \mu \frac{-2 b g_{\mathrm{Dn}} + b g_{\mathrm U} + 2 c g_{\mathrm{Dn}} - c h}
	{g_{\mathrm{Dn}} (b-c) \left[b (g_{\mathrm{Dn}} - g_{\mathrm U}) + c (h - g_{\mathrm{Dn}})\right]}.
\end{eqnarray}
These constitute the only two fixed points that possibly can lie within the simplex and is locally asymptotically stale. We mention here the other fixed points are either unstable or lie outside the simplex.

Using these expressions of fixed points, we compute the Jacobian matrix in terms of the parameters $b$, $c$, $h$, $g_{\mathrm U}$, and $g_{\mathrm{Dn}}$ to determine the parameter ranges for which the fixed points are stable. We employ \textit{Mathematica}~\cite{Mathematica} to symbolically evaluate the Jacobian and numerically compute the corresponding eigenvalues. Specifically, we discretize the parameter space by partitioning $h$, $g_{\mathrm U}$, and $g_{\mathrm{Dn}} \in [0,1]$ into 100 equally spaced values, while fixing $b=1$ and $c=0.1$. Substituting these grid points into the analytical expressions for the fixed points and eigenvalues allows us to systematically identify the regions of parameter space where the equilibria lie within the simplex and exhibit stability.

This choice of benefit ($b=1$) and cost ($c=0.1$) establishes a low cost-to-benefit ratio of $c/b=0.1$~\cite{Nowak1998a,Nowak1998b}. Such a low-cost regime reflects the empirical and theoretical reality that under indirect reciprocity frameworks, cooperative actions typically involve a modest immediate sacrifice by the donor while generating a substantially larger return for the recipient. In downstream reciprocity, cooperation is sustained via reputation building, where individuals incur a small immediate cost in exchange for the prospect of future benefits mediated by an enhanced public image. Similarly, in upstream reciprocity, passing a helping action forward is triggered by an internal state of gratitude, where the act itself usually imposes only a small immediate cost on the actor. Thus, assuming a low cost of cooperation provides a natural ground for studying indirect reciprocity. Moreover, because the evolutionary dynamics depend primarily on the cost-to-benefit ratio, fixing $b=1$ serves as a convenient normalization without any loss of generality. Therefore in this paper we will the numerical values $b=1$ and $c=0.1$.

The resulting stability regions of the parameters ($h$, $g_{\mathrm U}$, and $g_{\mathrm{Dn}}$) are illustrated in Fig.~\ref{fig:replicator_mutator_model}. As promised at the beginning of this section, our analysis confirms the existence of three prominent evolutionary outcomes: (i) a stable internal equilibrium corresponding to the coexistence of all three behavioural types, (ii) the dominance of downstream reciprocators, and (iii) a region of bistability between the coexisting state and the downstream-dominated state.

Above analysis completes our preliminary analytical results.
In the rest of the paper, we analyze the finite population model to explore the evolutionary dynamics among upstream reciprocators, downstream reciprocators, and defectors.
To capture the effect of population structure in the simplest possible way, we study the model on a random regular network~\cite{Lieberman2005,Newman2003,barabasi2016network} by varying the degree. Note that as the degree approaches the total number of nodes, the system converges to a well-mixed population; meaning, our analysis also encompasses the well-mixed population as a limiting case. Most importantly, we examine the role of the update rule and demonstrate that evolutionary outcomes are strongly sensitive to this choice.

\section{The conceptual framework}\label{sec:concept_framework}
Consider a finite population of $N$ individuals situated on a random regular network of degree $k$, where each node represents an agent belonging to one of three behavioural types: an upstream reciprocator, a downstream reciprocator, or a defector. At any instant, each individual can be in one of two roles: donor or receiver. The donor may choose to cooperate by helping the receiver at a cost $c$, thereby providing a benefit $b$ to the receiver, where $b > c$. If the donor defects, both individuals receive a payoff of zero. 

\begin{figure*}
	\includegraphics[scale=1.5]{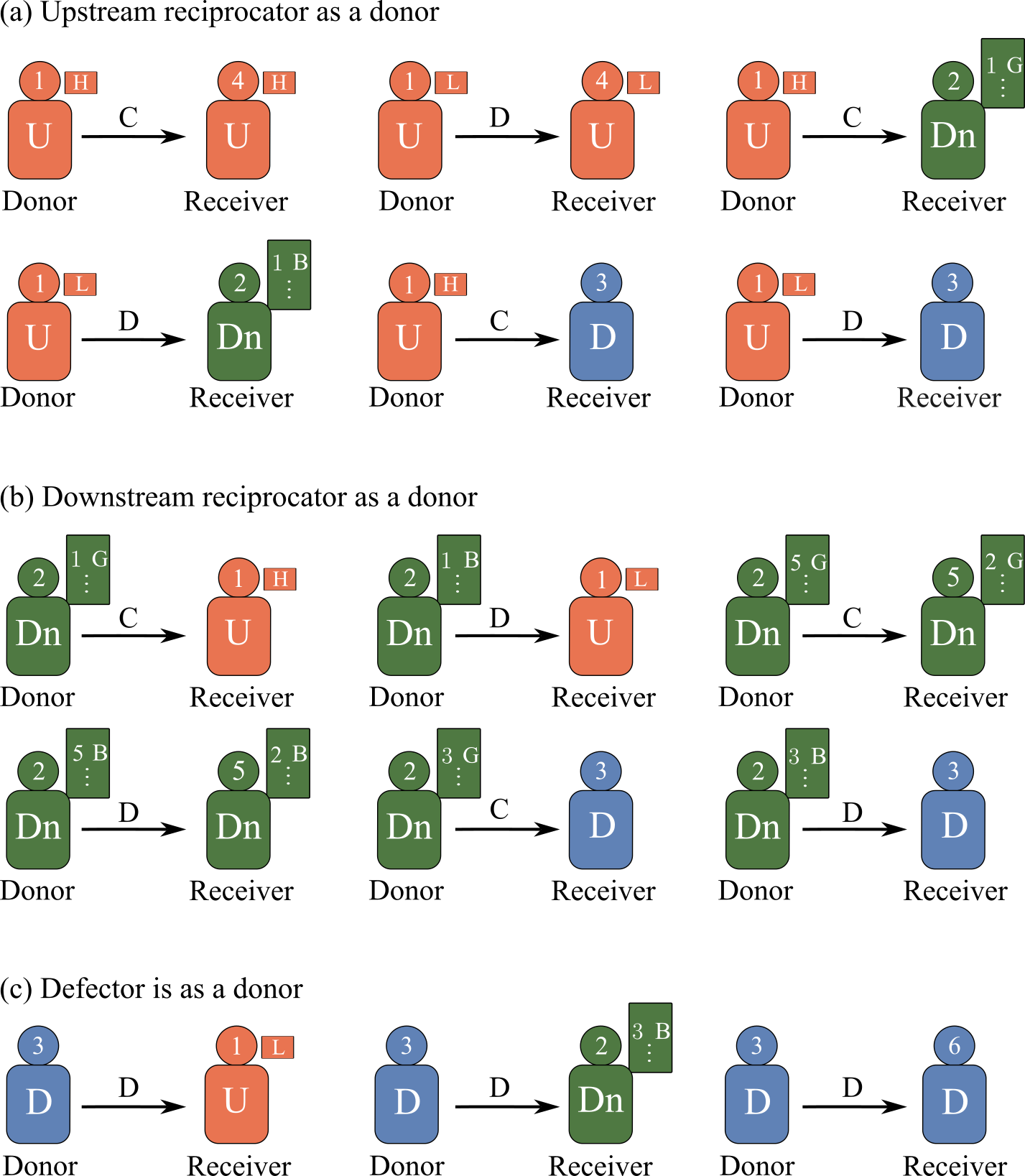}
	\caption{Schematic illustration of microscopic state updates across different pairwise interactions among the three behavioural types---upstream reciprocators (U), downstream reciprocators (Dn), and defectors (D). Each agent is assigned an identifier number (indicated on the head), and its behavioural type (U, Dn or D) is marked on the body. In each interaction block, the player on the left acts as the donor, whose action [either cooperation (C) or defection (D)] is indicated above the arrow, and the player on the right acts as the receiver. Each interaction dynamically alter individual microscopic state variables: public image-scores ($I_i \in \{{\rm G},{\rm B}\}$) and internal gratitude levels ($E_i \in \{{\rm H},{\rm L}\}$). For clarity, the green colored image-score box---adjacent to downstream donor---solely display the recipient's image (G or B along with the recipient identifier number) prior to the interaction. The image-score box adjacent to the downstream receiver only displays the donor's updated image after the interaction. Analogously, the red-colored gratitude level box---adjacent to upstream donor---display the donor's own internal gratitude state (H or L) prior to the interaction. The gratitude level box adjacent to the upstream receiver displays the receiver's own updated gratitude state after the interaction.}
	\label{fig:model}
\end{figure*}

Players perform their actions based on their behavioural type: Upstream reciprocators condition their behavior on their own internal emotional state representing gratitude. Each upstream reciprocator $i$ possesses a state variable $E_i \in \{{\rm H}, {\rm L}\}$, denoting high or low gratitude: When an upstream individual receives cooperation as a recipient, their state is updated to $E_i = {\rm H}$; conversely, when they experience defection from a donor, their state is set to $E_i = {\rm L}$. When subsequently acting as a donor, an upstream individual cooperates if $E_i = {\rm H}$ and defects if $E_i = {\rm L}$ as schematically illustrated in Fig.~\ref{fig:model}(a).

Downstream reciprocators condition their behavior on the image-scores of other individuals (whether they are upstream reciprocators, downstream reciprocators or defectors) which is being stored by each of the downstream reciprocator. Each individual $j$ in the population is assigned a binary image-score $I_j \in \{{\rm G}, {\rm B}\}$, corresponding to a "good" (G) or "bad" (B). Whenever a player's image changes, the update is publicly broadcast and observed by all individuals, leading to a consensus of opinion across the population. This common opinion can also be achieved by assuming that players gossip truthfully about the actions of others~\cite{Ohtsuki2004}. A downstream donor cooperates if the receiver has a good image ($I_j = {\rm G}$) and defects otherwise. Following the interaction, the donor's own image-score is updated according to their action: cooperation yields a good image ($I_i = {\rm G}$), while defection results in a bad image ($I_i = {\rm B}$). The actions (cooperate or defect) and corresponding update of internal states are pictorially demonstrated in Fig.~\ref{fig:model}(b). 
Note that in the previous section, we set $g_{\rm D}=0$, which translates to $I_j={\rm B}$ for any defector $j$. This assumption is relaxed here; because our agent-based simulations keep track of microscopic details, a defector who has not yet interacted may be perceived as good.
Finally, defectors unconditionally defect whenever they act as a donor, as shown in Fig.~\ref{fig:model}(c).

Depending on the scenario under consideration, we either restrict the population to a two-type setup---consisting solely of upstream reciprocators and defectors---or evaluate the full three-type configuration to systematically understand how the presence of downstream reciprocators influences the emergence of upstream reciprocity. We call the scenario with only upstream reciprocators and defectors as the \emph{baseline scenario}.

Having defined the behavioural types and how their internal state changes, we now describe the temporal structure of the interaction and reproduction processes. We explicitly distinguish between an evolutionary generation and rounds. Each evolutionary \emph{generation} consists of $R$ \emph{rounds} of interactions. In each round, a pair of connected individuals is randomly selected from the population, where one acts as the donor and the other as the receiver. The donor takes an action according to its behavioural type and current internal state as described in Fig.~\ref{fig:model}. At the end of each generation---which means upon the completion of all $R$ rounds---individuals reproduce proportionally to their accumulated fitness. To maintain a constant population size throughout the evolutionary process, newly born offspring replace existing individuals selected for removal according to the specific update mechanisms discussed below. Offspring inherit the behavioural type of their parent subject to a small mutation probability $\mu$, which ensures that all strategic states remain accessible and renders the underlying stochastic evolutionary process ergodic.

At this point, a critical question arises: How should newly born individuals perceive the surrounding population? Should they assume by default that all individuals possess a good reputation, or should they inherit the reputation structure already established in the population? In much of the existing indirect reciprocity literature where reputations are publicly accessible, it is a common practice to reset the public reputations of all individuals to a favorable state at the beginning of each generation~\cite{Hilbe2018,Ohtsuki2004}.

In systems with non-overlapping generations---where the entire population is replaced simultaneously and no surviving agents remain to carry over historical records---this question does not arise. Under such conditions, it is entirely logical to follow the traditional resetting convention, which assumes that newly born individuals perceive all members of the population as uniformly good by default, thereby resetting the reputations and gratitude of the entire population to G and H, respectively.

However, if generations are overlapping—where only a fraction of the population undergoes turnover via birth–death events while others persist—the choice of an intergenerational reset protocol is less straightforward. Here, two distinct possibilities emerge. The first approach is to uniformly reset the image-score and gratitude level of all individuals to G and H at the start of every generation. This protocol is widely adopted in the literature~\cite{Hilbe2018}, resting on the rationale that newborn individuals should enter a system where reputations are untainted. Under a strictly public information framework, maintaining this perceptual tabula rasa for newborns requires resetting the public scores of the entire collective. While introducing localized, private cognitive representations of image-scores for each agent could bypass this global reset, such an approach is computationally intensive, and we leave it for future work.

Nevertheless, one may fundamentally question why the behavioural histories of persisting individuals should be wiped clean simply because a few births occurred. Motivated by this query, we introduce and explore a second reinitialization protocol: resetting the image-score and gratitude level only for newly born individuals, while leaving the historical states of persisting agents intact. This realistic condition has received minimal attention in existing literature, and we demonstrate that retaining these historical states can substantially alter the macroscopic evolutionary outcomes.

\begin{figure}[h]
	\includegraphics[scale=0.5]{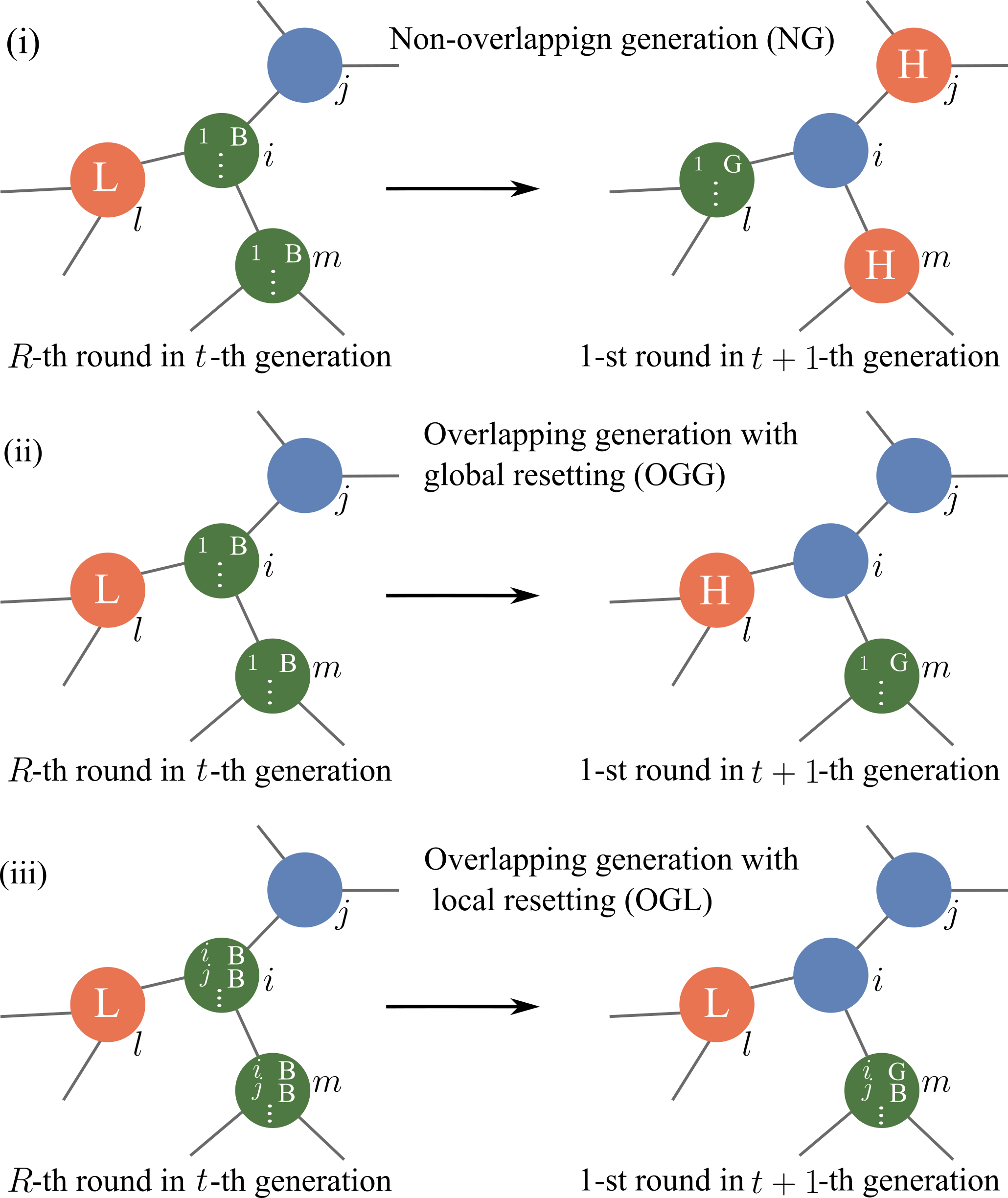}
	\caption{
		Schematic representation of the three generational configurations evaluated in this study. Representative network nodes denote individuals belonging to one of the three behavioural types: upstream reciprocators (red nodes), downstream reciprocators (green nodes), and defectors (blue nodes). The state transitions are shown from the final ($R$-th) round of generation $t$ to the initial ($1$-st) round of generation $t+1$.
		(i) Non-overlapping generations (NG): The entire population is replaced at the generational boundary. Consequently, node behavioural types can completely change, image-scores are universally reset to G (as shown in the registry entry for the new node $l$), and internal gratitude levels are uniformly reset to H (as shown for the new nodes $j$ and $m$). (ii) Overlapping generations with global resetting (OGG): Only a single agent (node $i$) is replaced by a newborn, yet the states of all individuals are globally reinitialized. Thus, the image-score of the persisting node $m$ is reset to G, and the internal gratitude level of the persisting node $l$ is reset to H. (iii) Overlapping generations with local resetting (OGL): Only the newborn agent (node $i$) has its public image-score initialized to G, while the historical states of all persisting agents remain completely unaltered. In contrast to OGG, the internal gratitude level of the persisting node $l$ remains at L, and the image entry of the persisting node $m$ retains its historic state, while only the public entry for the newborn node $i$ is initialized to G.}
	\label{fig:schematic}
\end{figure}

Therefore, in this paper, we systematically investigate three distinct generational configurations  (see Fig.~\ref{fig:schematic}):
\begin{enumerate}[(i)]
	\item \textbf{NG:} \textbf{N}on-overlapping \textbf{G}enerations, where the entire population is simultaneously replaced at the end of each generation.
	\item \textbf{OGG:} \textbf{O}verlapping \textbf{G}enerations with \textbf{G}lobal resetting, where the public image-scores and internal gratitude levels of all individuals are reset to G and H, respectively, at the start of every generation.
	\item \textbf{OGL:} \textbf{O}verlapping \textbf{G}enerations with \textbf{L}ocal resetting, where only newly born individuals have their states reinitialized to G and H upon entry, while the historical states of persisting individuals remain unaltered.
\end{enumerate}

These three generational update mechanisms are implemented alongside their corresponding overlapping and non-overlapping reproductive frameworks, the exact algorithmic details of which are described in the next section.

\section{Numerical scheme}\label{sec:numercial_scheme}
We now elaborate on the specific protocols governing our numerical investigations, outlining the precise computational architecture and the key modeling assumptions embedded within our simulation framework.

We start our simulations on a random regular network consisting of $N$ nodes, where each node has a fixed degree $k$. Every node represents an individual belonging to one of the three behavioural types: upstream reciprocators, downstream reciprocators, or defectors. The individuals interact pairwise over $R$ successive rounds within each generation $t$ according to the following protocol: In each round, a pair of connected nodes (neighbors) is selected uniformly at random, with one designated as the donor and the other as the receiver. The donor decides whether to cooperate---incurring a minimal cost $c$ to provide a benefit $b$ to the receiver---or defect to pay a zero cost and provide zero benefit, based strictly on its behavioural type and current internal state (Fig.~\ref{fig:model}). Following the interaction, all downstream reciprocators synchronously update their records of the donor's image-score, where cooperation yields a good image (G) and defection yields a bad image (B). Conversely, an upstream individual updates its gratitude state only when acting as the receiver, shifting to high gratitude (H) if the donor cooperated and to low gratitude (L) if the donor defected. 

This interaction and state-update process is repeated for a total of $R$ rounds, during which individuals accumulate their respective payoffs. Since payoffs are frequency-dependent and therefore determined by the current population composition, any change in composition due to reproduction modifies the payoff environment experienced by individuals. Accordingly, payoffs are accumulated only within the generation and are reset after reproduction, so that fitness comparisons in the next generation are based solely on interactions occurring under the updated population composition. This convention is consistent with evolutionary processes in which payoffs are computed for every composition~\cite{Hilbe2018,Schmid2021}. These interaction rounds remain same across all three update schemes discussed next; they differ only in the reproduction process that occurs at the end of each generation.

In the case of non-overlapping generations (NG), the entire population reproduces simultaneously in proportion to their accumulated payoffs while maintaining a constant population size ($N$). For each individual $i$ (the focal individual), we consider its accumulated payoff alongside the payoffs of its $k$ immediate neighbors. This local payoff landscape---comprising the payoff of individual $i$ and those of its $k$ neighbors---defines a \emph{local discrete probability distribution} over the $k+1$ individuals, where the probability of selecting any individual (from the $k+1$ choices) is directly proportional to their accumulated payoff. 
Under this framework, the focal individual $i$ randomly adopts the strategy of a neighbor $j$ sampled from this local discrete probability distribution. This strategy adoption is additionally subject to mutation: with a small probability $\mu$, individual $i$ mutates to any of the three behavioural types (U, Dn, or D) with equal probability~\cite{Nowak1998a}, whereas with probability $1-\mu$, the strategy is faithfully adopted from the sampled individual $j$.

This selection and mutation process is executed independently for all $N$ individuals in the population to form the next generation. Finally, as discussed at the end of Sec.~\ref{sec:concept_framework}, all image-scores and gratitudes are reset to G and H respectively at the beginning of each generation. 

In the case of overlapping generations with global resetting (OGG), the update protocol is substantially different from the non-overlapping (NG) case. In this scenario, upon the completion of $R$ rounds of interaction (completion of a generation), pairs of individual are selected from the population to update their strategies. The probability that a focal individual $i$ adopts the behavioural type of an neighboring individual $j$ is governed by the standard pairwise Fermi imitation rule~\cite{2007_Szabo_Fath,Blume1993,Traulsen2006}:
\begin{equation}\label{eqn:fermi_rule}
	P(i\to j)=\frac{1}{1+\exp[-\beta(\pi_j-\pi_i)]}.
\end{equation}
where $\pi_i$ and $\pi_j$ denote the accumulated payoffs of the $i$-th and $j$-th individuals over the $R$ interaction rounds, respectively, and $\beta\geq 0$ represents the intensity of selection. It is clear that this scheme also preserve the number of individuals $N$.

To make the simulations computationally viable, we introduce a slight modification to the standard sequential implementation of this update rule. If only a single pair of individuals is selected for payoff comparison and strategy updation in each generation, the system's relaxation time toward its stationary distribution becomes extremely slow—particularly in large populations—resulting in prohibitively long simulation runtimes. To overcome this computational bottleneck, instead of updating a single pair asynchronously, we implement a \emph{semi-synchronous} block update scheme where approximately 10\% of the population is updated in parallel during each generation. Beyond its computational advantages in accelerating numerical convergence, this semi-synchronous approach provides a more realistic representation of evolutionary and biological systems compared to strict serialization~\cite{Bach2003, wierniak2016}. As we have checked, this modification leaves the qualitative macro-level evolutionary dynamics entirely unchanged. Consequently, throughout our analysis of overlapping generations, we systematically employ this 10\% parallelized updating protocol.

In case of overlapping generation with local resetting (OGL) we introduce a more localized initialization scheme: only newly born individuals are assigned a good public reputation and a high gratitude state ($I_i={\rm G}$ and $E_i = {\rm H}$), while the remaining, surviving individuals retain their pre-existing internal states. For the evolutionary dynamics, we employ the identical pairwise Fermi imitation rule [Eq.~(\ref{eqn:fermi_rule})] to determine the probability that a randomly selected individual $i$ adopts the behavioural type of randomly selected neighbor $j$. As in the previous OGG framework, to accelerate numerical convergence we apply the semi-synchronous updating approach in each generation~\cite{Bach2003, wierniak2016}.

We simulate these three evolutionary process separately over a large number of generations in Sec.~\ref{sec:Results} and compute the time-averaged frequencies to approximate the steady state frequency distribution of the three types, $x_{\rm U}$, $x_{\rm Dn}$, and $x_{\rm D}$, defined as:
\begin{equation}\label{eqn:steady_state}
	x_{\rm U} = \frac{1}{T} \sum_{t=1}^T \frac{n_{\rm U}(t)}{N},~
	x_{\rm Dn} = \frac{1}{T} \sum_{t=1}^T \frac{n_{\rm Dn}(t)}{N}~\text{and}~x_{\rm D} = \frac{1}{T} \sum_{t=1}^T \frac{n_{\rm D}(t)}{N}.
\end{equation}
Here, $n_{\rm U}(t)$, $n_{\rm Dn}(t)$, and $n_{\rm D}(t)$ denote the number of upstream reciprocators, downstream reciprocators, and defectors in $t$-th generation, respectively, while $T$ represents the total number of generations over which the simulation average is taken. 

To further characterize the system, we also compute the total cooperation level $C_{\rm T}$, defined as the time-averaged fraction of cooperative actions across the population:
\begin{equation}\label{eqn:cooperation_level}
	C_{\rm T} = \frac{1}{T} \sum_{t=1}^T \frac{c(t)}{R},
\end{equation}
where $c(t)$ denotes the total number of cooperative actions executed by donors during the $R$ interaction rounds of the $t$-th generation.

At this stage, it is appropriate to introduce a key parameter: the average number of interactions per pair, which dictates how frequently a specific dyad is selected to interact within a single generation. When this quantity is small, a given pair is selected, on average, only once, minimizing the likelihood of direct reciprocity. We define this parameter, denoted by $\eta$, as the ratio of the total number of interaction rounds ($R$) per generation to the number of distinct edges (number of connected pairs, $Nk/2$) in the network:
\begin{equation}\label{eqn:eta}
	\eta = \frac{2R}{Nk}.
\end{equation}
We refer to $\eta$ as the average sampling frequency per pair. A high value of $\eta$ implies that the exact same pair is selected multiple times during a generation. In the well-mixed limiting case where $k = N - 1$, this scales as $\eta = \frac{2R}{N(N-1)}$. Crucially, to isolate the effects of indirect reciprocity, we fix $\eta = 1$ throughout this paper. Ensuring that each connected pair is sampled on average only once per generation automatically binds the interaction rounds to the network topology via the relation $R = Nk/2$.

\section{Results}\label{sec:Results}
With the model setup established in Secs.~\ref{sec:concept_framework} and \ref{sec:numercial_scheme}, we now investigate the evolutionary dynamics of upstream reciprocators, downstream reciprocators, and defectors under each of the three generational configurations illustrated in Fig.~\ref{fig:schematic}. Across our analysis, we first evaluate a baseline scenario involving only upstream reciprocators and defectors, explicitly highlighting the sensitivity of upstream reciprocity to exploitation by defectors. We then introduce downstream reciprocators into the population to examine how a public reputation-based mechanism alters these evolutionary outcomes. In particular, we focus on how the steady-state frequency of upstream reciprocators [Eq.~(\ref{eqn:steady_state})] shifts in the presence of downstream reciprocators, and how their interaction influences the macroscopic cooperation level [Eq.~(\ref{eqn:cooperation_level})]. In all the analysis we are going to systematically map out how the steady-state frequencies of the behavioural types vary across different combinations of the population size ($N$) and the network degree ($k$).

\subsection{Non-overlapping generation (NG)}\label{sec:NG}
\begin{figure*}
	\includegraphics[scale=1.0]{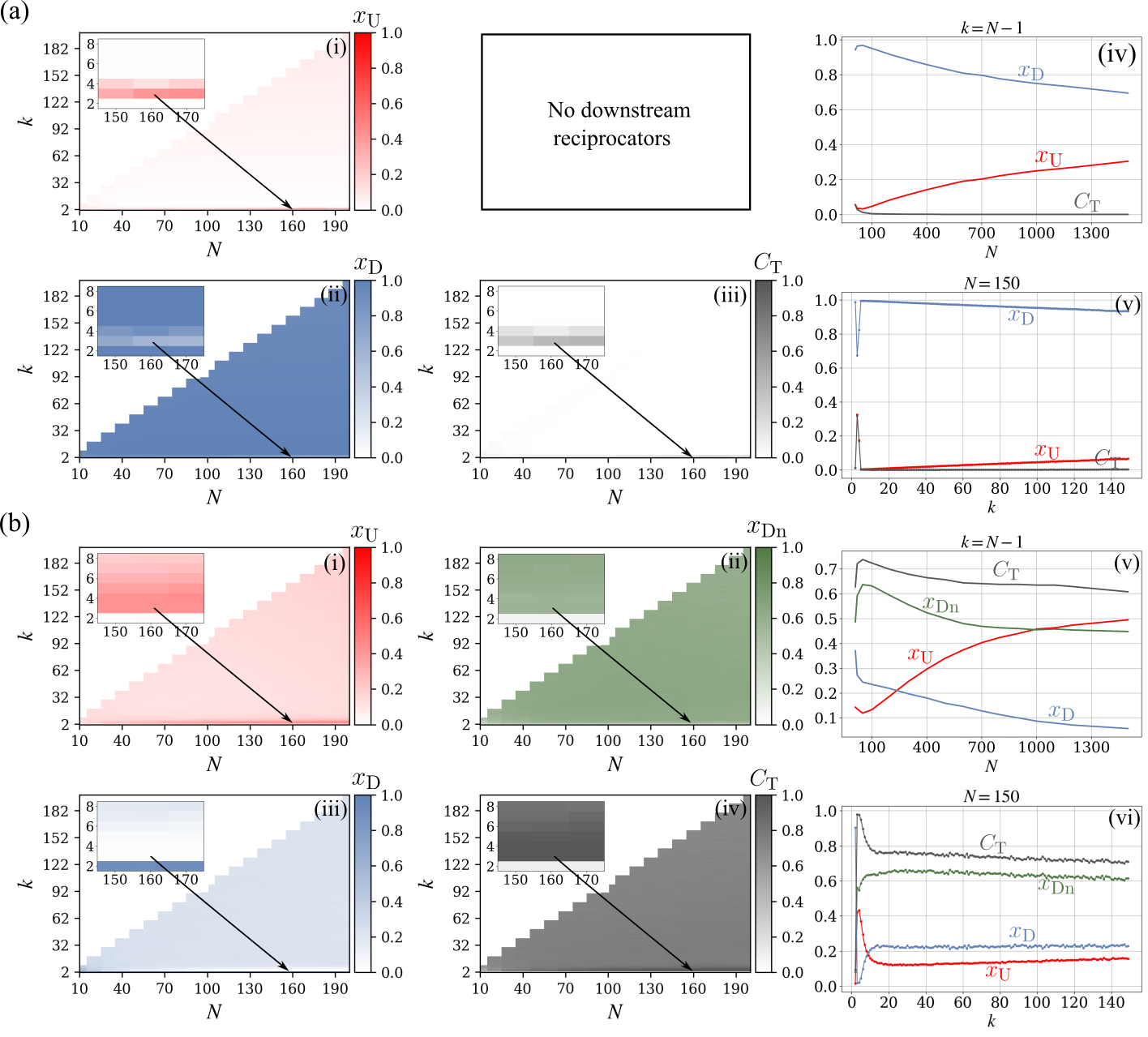}
	\caption{Steady-state frequency distributions and cooperation levels under the non-overlapping generations (NG) framework:  Results are mapped as a function of the population size $N$ and network degree $k$. Panel~(a) shows results in the absence of downstream reciprocators (baseline scenario), while panel~(b) shows results when all three behavioural types are present (full system). In both major blocks, the heatmaps display the long-term steady-state frequencies ($x_{\rm U}$, $x_{\rm Dn}$, $x_{\rm D}$) and cooperation level ($C_{\rm T}$) across the $(N, k)$ parameter space. The insets in (a)(i)--(iii) and (b)(i)--(iv) show magnified views of the small-$k$ region near the $N$ values indicated by the arrow. Panels~(a)(iv) and (b)(v) show the dependence of strategy frequencies on $N$ alone, with the degree fixed at $k = N-1$. Panels~(a)(v) and (b)(vi) show the dependence on $k$ alone, with population size fixed at $N = 150$.}
	\label{fig:NG}
\end{figure*}

We examine the steady-state frequencies of both the baseline scenario (comprising upstream reciprocators and defectors) and the full three-type system across the entire $(N,k)$ parameter space, the results of both the cases are presented in Fig.~\ref{fig:NG}.

\subsubsection*{Baseline scenario: In the absence of downstream reciprocators}
The first observation we make from our simulations is that defectors completely dominate the population, with only a small fraction of upstream reciprocators persisting across nearly the entire $(N,k)$ parameter space shown in Fig.~\ref{fig:NG}(a). Consequently, the overall cooperation level remains close to zero across virtually all population sizes and network degrees. Despite this global pattern of defector dominance, distinct localized trends emerge when the results are examined along two axes: varying $N$ while keeping $k$ fixed, and varying $k$ while keeping $N$ fixed [see Figs.~\ref{fig:NG}(a)(iv)–(v)].

Let us first examine the cross-sectional slices of Figs.~\ref{fig:NG}(a)(i)--(iii) where $N$ is varied while keeping the network degree fixed at the well-mixed limit, $k = N-1$, as shown in Fig.~\ref{fig:NG}(a)(iv). This particular choice allows us to directly contrast our finite-population results with the analytical benchmarks derived for infinite population in Sec.~\ref{sec:INF_model}. Notably, this qualitative pattern remains robust for other choices of $k$, as evidenced by the full $(N,k)$ parameter space plots in Figs.~\ref{fig:NG}(a)(i)–(iii). We observe that although defectors dominate the population globally, the steady-state frequency of upstream reciprocators monotonically increases with $N$ [Fig.~\ref{fig:NG}(a)(iv)], consequently the defector frequency decreases with $N$. Interestingly, this numerical increase in upstream reciprocators' frequency is completely decoupled from the overall cooperation level, which remains suppressed near zero. This difference uncovers a vital feature of this scenario: the upstream reciprocators that survive and expand in larger populations are predominantly trapped in the low gratitude state ($E_i = \rm{L}$). Consequently, they rarely initiate cooperation, leading to a persistently low macroscopic cooperation level.

Next, we turn to the cross-sectional slices of Figs.~\ref{fig:NG}(a)(i)--(iii) where the network degree $k$ is varied while keeping the population size fixed at $N = 150$. As illustrated in Fig.~\ref{fig:NG}(a)(v), the population remains heavily dominated by defectors across this entire regime as well. However, we observe a distinct non-monotonic dependence on $k$: the steady-state frequency of upstream reciprocators exhibits a pronounced local maximum at small network degrees, centered near $k = 3$. As $k$ increases past this point, their frequency initially drops before experiencing a gradual, modest recovery at higher degrees, which is accompanied by a corresponding decline in the defector fraction. Notably, the overall cooperation level also exhibits a localized peak near $k = 3$. Yet, in stark contrast to the frequency of upstream reciprocators, $C_{\rm T}$ remains completely suppressed near zero for all larger values of $k$.

These observations raise two important questions that need to be addressed:
\begin{enumerate}[(i)]
	\item Why does the frequency of upstream reciprocators increase with $N$.
	\item Why does the frequency of upstream reciprocators exhibit a local maximum at some value of $k$, which we denote by $k_{\max}$ (=3)?
\end{enumerate}

The mechanism underlying the first question can be understood through the balance between selection pressure and mutant influx. Under the NG framework, the entire population is replaced at each generational transition. During this reproduction phase, an offspring copies its parent's behavioural type with high fidelity, but undergoes a mutation event with a small probability $\mu$, adopting a random behavioural type. Because this microscopic reproduction process is executed $N$ times to reconstruct the entire population, the expected number of mutation events injected into the system per generation scales as $N\mu$. Consequently, when $N$ is small, this absolute  \emph{mutational influx} is exceptionally low, meaning the population spends prolonged periods in monomorphic states dominated by the evolutionary favored type. In contrast, as $N$ grows large, the absolute number of mutants introduced per generation increases proportionally.

Even though defectors maintain a systematic payoff advantage and are strictly favored by selection, the continuous influx of mutants ensures that upstream reciprocators are constantly reintroduced into the population. Once present, these individuals participate in subsequent interaction rounds and retain a nonzero probability of reproducing—proportional to their accumulated payoffs. While these mutants are often entirely absent in small populations due to stochastic extinction, their frequent reintroduction in larger populations sustains a higher, steady-state frequency.

This frequency elevation eventually saturates at very large values of $N$ because selection continues to penalize the upstream strategy. The frequency $x_{\rm U}$ thus plateaus at an asymptotic value dictated by the relative payoff differences between upstream reciprocators and defectors. Crucially, because interactions with defectors systematically drive the internal gratitude levels of these surviving upstream reciprocators to the low state ($E_i = {\rm L}$), this inflation in $x_{\rm U}$ fails to manifest as an increase in the global cooperation level $C_{\rm T}$. The vast majority of persisting upstream reciprocators remain trapped in the ungrateful state.

The mechanism driving the second phenomenon, namely the emergence of $k_{\max}$, stems from a tension between two distinct network interactions: the local clustering of upstream reciprocators, which insulates them from exploitation, and their interface with defectors, which scales with the number of shared boundary edges. While the former process promotes the persistence and viability of the upstream reciprocators, the latter systematically erodes them. The non-monotonic interplay between these opposing forces gives rise to an optimal window. We formally quantify this trade-off in Appendix~\ref{sec:kmax_network} by calculating the average cluster sizes and the average number of shared boundaries, confirming that this evolutionary sweet spot occurs precisely at $k_{\max}=3$. The intuitive understanding of this behavior can be unpacked as follows.

At $k=2$ (a one-dimensional ring topology), the frequency of upstream reciprocators is severely suppressed. This is because the NG framework replaces the entire population, hence the distribution of behavioural types changes rapidly from one generation to the next, making clusters of upstream reciprocators difficult to sustain. Furthermore, even when micro-clusters of upstream reciprocators occasionally coalesce, they are necessarily surrounded by defectors from both sides along the ring. This maximizes their exposure to exploitation, which quickly drives their internal gratitude levels to the ungrateful state ($E_i = {\rm L}$). Consequently, their average payoff drops, rendering the upstream reciprocators evolutionarily non-viable at $k=2$.

This situation is significantly changes at $k=3$. At this higher degree, the increased local branching allows upstream reciprocators a higher probability of forming clusters. These spatial structures act as relatively insulated safe havens where upstream reciprocators can frequently cooperate with one another, bolstering their local payoffs and reproductive success. Although boundary interactions with defectors still occur, internal interactions within the core of the cluster frequently replenish gratitude levels, sustaining higher emotional gratitudes and lifting the cooperation level. While defectors also form spatial clusters at $k=3$, such structures offer them no evolutionary benefit because mutual defection yields zero payoff. Defectors remain entirely dependent on skimming payoffs from the edges shared with upstream reciprocators. When the network degree is increased beyond $k=3$, the network approaches a more well-mixed state. This density increase drastically multiplies the number of paths through which defectors can penetrate the clusters, breaking down the spatial insulation and exposing the upstream reciprocators to exploitation. Consequently, the beneficial effects of clustering are rapidly watered down, explaining why both the frequency of upstream reciprocators and the cooperation level peak sharply at $k_{\max}=3$ before collapsing at higher degree.

Interestingly, far beyond $k_{\max}$, the frequency of upstream reciprocators exhibits a slight upward trend at high network degrees. However, this recovery is entirely absent in the cooperation level. This distinction between the two trends suggests that in highly interconnected networks, upstream reciprocators interact more frequently with defectors, which drives them into the ungrateful state ($E_i={\rm L}$). As a result, these upstream reciprocators rarely cooperate, which minimizes their exploitation. Under these conditions, the payoffs of upstream reciprocators and defectors becomes comparable because both behavioural types effectively function as defectors. This narrowing of the payoff deficit relaxes the selective pressure against the upstream strategy, allowing its frequency to drift upward. Furthermore, this recovery becomes significantly more pronounced in larger populations due to the enhanced mutational influx ($N\mu$) discussed previously.

\subsubsection*{Full system: In the presence of downstream reciprocators}

After establishing the baseline scenario, we now introduce downstream reciprocators into the population to evaluate the full three-type system. The immediate consequence—visible across the entire $(N,k)$ parameter space in Fig.~\ref{fig:NG}(b)—is a substantial increase in the steady-state frequency of upstream reciprocators, accompanied by a corresponding collapse in the defector frequency. Furthermore, the global cooperation level rises significantly, a shift that becomes immediately apparent when comparing the baseline heatmap in Fig.~\ref{fig:NG}(a)(iii) with the heatmap in Fig.~\ref{fig:NG}(b)(iv). While cooperative behavior is virtually absent in the baseline scenario, the inclusion of reputation-based downstream reciprocators fosters a markedly more cooperative environment. Beyond these global patterns, several intriguing evolutionary phenomena emerge when we explore the data along two specific parametric directions: varying $N$ while keeping $k$ fixed, and varying $k$ while keeping $N$ fixed.

We first explore the cross-sectional slices of Figs.~\ref{fig:NG}(b)(i)--(iv) where the population size $N$ is varied while keeping the network degree fixed at the well-mixed limit ($k = N-1$), as shown in Fig.~\ref{fig:NG}(b)(v). As $N$ increases, the steady-state frequency of upstream reciprocators monotonically increases and eventually saturates at a significantly higher value compared to the baseline case. Concurrently, the frequency of defectors drops to a heavily suppressed level, while the fraction of downstream reciprocators stabilizes at a finite value. Consequently, the system settles into a state of effective three-type coexistence. This finite-population behavior elegantly matches with mean-field predictions derived for the infinite-population limit in Sec.~\ref{sec:INF_model}, where stable three-type (U, Dn, and D) coexistence is analytically supported.

Next, we consider the corresponding slices of Figs.~\ref{fig:NG}(b)(i)--(iv) where the network degree $k$ is varied while keeping the population size fixed at $N = 150$, as illustrated in Fig.~\ref{fig:NG}(b)(vi). Here, the frequency of upstream reciprocators is severely minimized at ring limit of $k = 2$. It then climbs rapidly with increasing connectivity, reaching a distinct local maximum around $k_{\max}=4$, before decaying and saturating at a lower value in denser networks. In contrast, the frequency of downstream reciprocators starts at a lower value at $k=2$ and monotonically increases, saturating at a higher value without exhibiting a local peak around any intermediate $k$. Notably, this saturation value of the upstream strategy remains substantially higher than the baseline observed in the absence of downstream reciprocators. Furthermore, while the absolute magnitude of this saturation plateau scales upward with larger population sizes ($N$), the overall qualitative, non-monotonic dependence on $k$ remains unchanged.

These rich observations naturally demand a microscopic explanation for two key phenomena:
\begin{enumerate}[(i)]
	\item Why the steady-state frequency of upstream reciprocators is systematically amplified by increasing $N$, and 
	\item Why the local maximum of upstream frequency occur at $k_{\max}$ ($=4$)?
\end{enumerate}

The mechanism underlying the first question can be mapped to the reputation-mediated dynamics of internal gratitude states. Under the NG framework, the internal gratitude levels of all upstream individuals are refreshed at the beginning of each generation, enabling them to initially cooperate. Consequently, these upstream actors systematically acquire a good public reputation ($I_i = {\rm G}$), whereas defectors rapidly accumulate a bad reputation ($I_i = {\rm B}$) due to repeated exploitation of their partners. Downstream reciprocators strongly reinforces this behavioural asymmetry: They preferentially reward upstream individuals holding a high gratitude state ($E_i = {\rm H}$) while strictly withholding cooperative benefits from defectors. Because upstream reciprocators pay it forward the received help, a powerful positive \emph{feedback loop} emerges: An upstream donor who receives help is driven into the high gratitude state, increasing their likelihood of subsequent cooperation and sustaining a pristine public reputation, which in turn invites further cooperation from downstream reciprocators.

This enhancing effect intensifies with increasing population size because the influx of mutant strategies scales linearly as $N\mu$. In this regime, the continuous injection of downstream mutants plays a critical role. By effectively distinguishing between the favorable reputations of upstream reciprocators and the poor reputations of defectors, these downstream actors preferentially channel payoffs toward the former. As $N$ expands, the absolute abundance of downstream reciprocators increases, causing upstream individuals to receive a disproportionally higher share of the payoff relative to defectors. Although this mutation-driven inflation eventually saturates at very large $N$, the asymptotic saturation level of $x_{\rm U}$ remains remarkably high. Ultimately, downstream reciprocators foster the evolution of upstream reciprocators: they systematically suppress the payoff of defectors while bolstering the fitness of upstream reciprocators, allowing upstream reciprocity to thrive in large populations and catalyze cooperation.

Notably, the convex curvature observed in the upstream frequency at smaller $N$ [Fig.~\ref{fig:NG}(b)(v)] reflects the well-known small-group effect, which inherently favors localized upstream reciprocity~\cite{Pfeiffer2005,Boyd1989}. Crucially, our results demonstrate that via this reputation-based feedback loop, upstream reciprocity can robustly persist even when scaling into the large $N$ limit. 

The second question regarding the emergence of $k_{\max}=4$ [see Fig.~\ref{fig:NG}(b)(vi)] is particularly compelling, as it highlights how downstream reciprocators shift the peak of the upstream frequency compared to the baseline case. Four distinct network factors govern the evolutionary dynamics here: (i) the upstream clusters, (ii) the downstream clusters, (iii) the density of the edges between upstream and downstream reciprocators, and (iv) the density of edges shared between upstream reciprocators and defectors. Based on what we have discussed so far, it is easy to deduce that the first three factors exert a positive influence on the fitness of upstream reciprocators, whereas the final factor imposes negative impact. While the quantitative trade-offs among these four competing factors are formulated in Appendix~\ref{sec:kmax_network}, the underlying intuition can be unpacked as follows.

At $k=2$ (one-dimensional ring topology), defectors strongly hinders the propagation of cooperative behavior. Because each individual possesses only two neighbors, any local cluster of upstream or downstream individuals is easily blocked by defectors. In particular, downstream reciprocators at this degree find themselves isolated and interacting predominantly with defectors; unable to find cooperative rewards, strictly limiting the propagation of reciprocity. However, when $k$ increases from $2$ to $3$, the expanding local branching factor grants individuals additional neighbors, drastically increasing the likelihood of cross-links forming between upstream and downstream reciprocators. This structural shift facilitates the cooperative clusters and a simultaneous surge in the frequencies of both reciprocating types. As the network degree increases to $k=4$, this mutual clustering is optimized, maximizing their collective payoff advantage and enabling the cooperative alliance to outperform defectors and achieve peak frequencies.

Beyond $k_{\max}=4$, entering the intermediate-to-high connectivity regime ($k \ge 5$) renders the population more well-mixed, exposing both cooperative types to a higher density of defectors. This structural dissolution penalizes upstream reciprocators far more severely than downstream reciprocators. Because downstream reciprocators can actively discriminate against defectors based on defectors' poor public reputation, they selectively channel benefits toward cooperative neighbors. Upstream reciprocators, conversely, lack this reputational shield leaving them vulnerable to exploitation. Consequently, the frequency of upstream reciprocators begins to decline past $k_{\max}=4$, while the downstream strategy—shielded by its reputational filtering---saturates.

This decline eventually stabilizes for sufficiently large values of $k$. As the network approaches the well-mixed limit, the localized encounter rates between upstream reciprocators, downstream reciprocators, and defectors become approximately constant. As a result, the frequency of upstream reciprocators flattens into a saturation plateau. Notably, this plateau remains substantially higher than the baseline observed in the absence of downstream reciprocators, demonstrating the protection offered by the reputation mechanism.

Finally, this saturation plateau scales upward with the population size $N$. As the population size grows, the mutational influx ($N\mu$) intensifies. This heightened influx ensures that downstream reciprocators repeatedly emerge to undermine defector-dominated clusters, while upstream reciprocators can smoothly proliferate within the hospitable, reputation-cleansed environments sustained by downstream individuals. This evolutionary synergy ultimately amplifies the background frequency of upstream reciprocators as $N$ increases.

\subsection{Overlapping generation with global resetting (OGG): All the individuals' image-score and gratitude reset in every generation}\label{sec:overlapping_all_Update}
\begin{figure*}
	\includegraphics[scale=1.0]{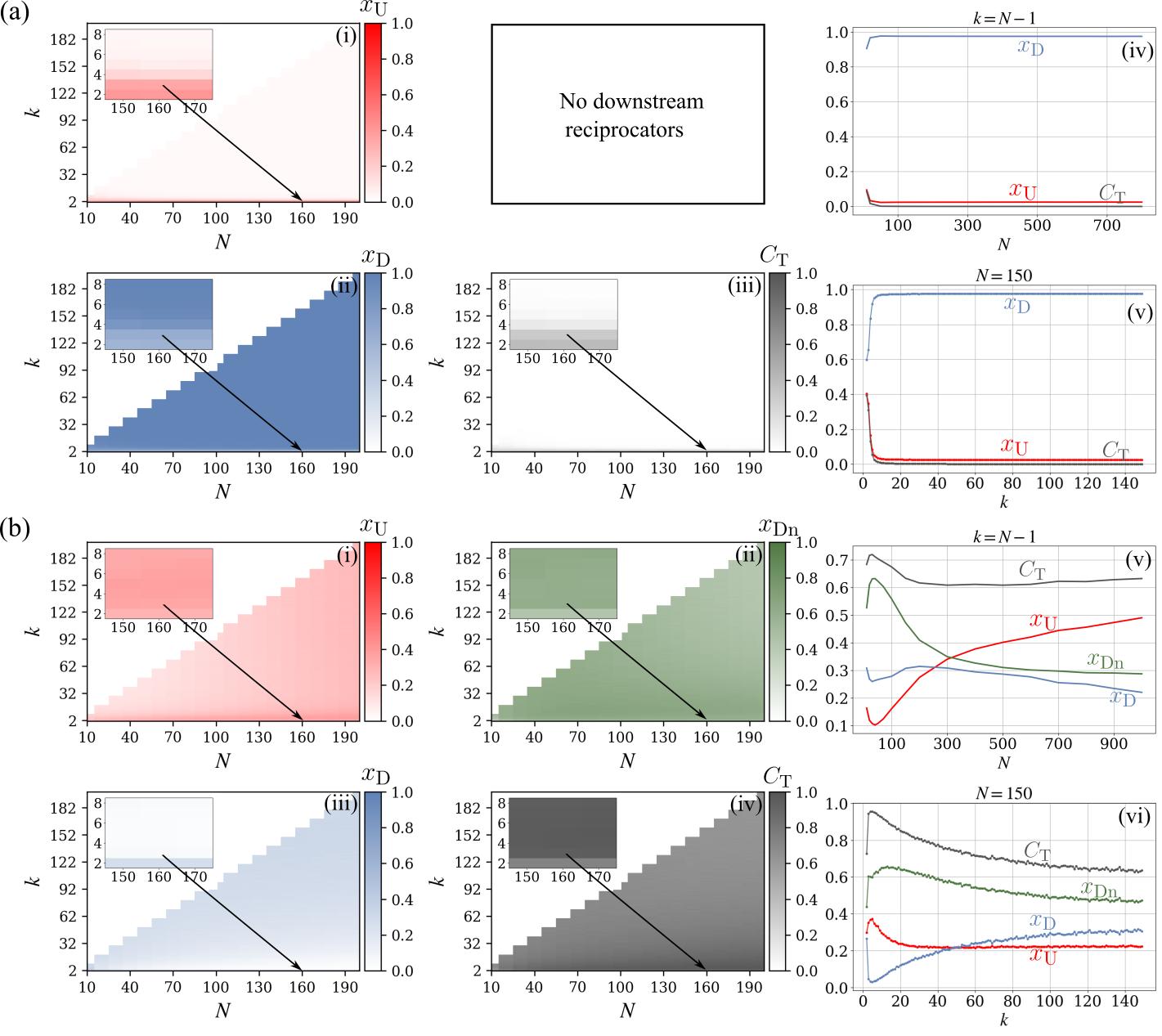}
	\caption{Steady-state frequency distributions and cooperation level under the overlapping generation with global resetting (OGG) framework: Results are mapped as a function of the population size $N$ and network degree $k$. Panel~(a) shows the outcomes in the absence of downstream reciprocators (baseline scenario), while panel~(b) shows results when all three behavioural types are present (full system). In both major blocks, the heatmaps display the long-term steady-state frequencies ($x_{\rm U}$, $x_{\rm Dn}$, $x_{\rm D}$) and cooperation level ($C_{\rm T}$) across the $(N, k)$ parameter space. The insets in (a)(i)--(iii) and (b)(i)--(iv) demonstrate magnified views of the small-$k$ region near the $N$ value indicated by the arrow. Panels~(a)(iv) and (b)(v) show the dependence of strategy frequencies on $N$ alone, with the degree fixed at $k = N-1$. Panels~(a)(v) and (b)(vi) show the dependence on $k$ alone, with population size fixed at $N= 150$.}
	\label{fig:overlapping_population_all_update}
\end{figure*}

As mentioned at the beginning of this section, the public reputations and internal gratitude states of all individuals are refreshed in every generation. As before, to investigate the dependence of the steady-state frequencies of the three behavioural types on $N$ and $k$, we plot the stationary distributions across the $(N,k)$ parameter space in Fig.~\ref{fig:overlapping_population_all_update}.

\subsubsection*{Baseline scenario: In the absence of downstream reciprocators}

We again begin our investigation with the baseline scenario where only upstream reciprocators and defectors are present. As illustrated in Fig.~\ref{fig:overlapping_population_all_update}(a), defectors continue to dominate the population. However, in stark contrast to the NG framework, the stationary distributions across the $(N,k)$ parameter space exhibit a different qualitative pattern, highlighting how sensitively these evolutionary outcomes depend on the underlying microscopic update rule. To gain deeper insight into these variations, we systematically analyze the results along the two cross-sectional directions in Figs.~\ref{fig:overlapping_population_all_update}(a)(i)--(iii): varying $N$ while keeping $k$ fixed, and varying $k$ while keeping $N$ fixed.

We first consider the cross-section at the well-mixed limit ($k=N-1$) and vary the population size $N$. As shown in Fig.~\ref{fig:overlapping_population_all_update}(a)(iv), we find that the steady-state frequency of upstream reciprocators remains suppressed near zero and exhibits negligible dependence on $N$, whereas the defector fraction stays consistently close to 100\%.
This result is fully consistent with the infinite-population analysis in Sec.~\ref{sec:INF_model}, where only the fixed point corresponding to complete defection is stable in the reduced system [Eq.~(\ref{eq:reduced_sys})].
The reason for this absence of frequency expansion with increasing $N$---which stands in stark contrast to the NG framework---can be understood through the mechanics of OGG. Under OGG update rule, only a small fraction of the population modifies their behavioural type in each generation, unlike the NG framework where the entire population is replaced simultaneously. Consequently, the generation-wide mutational influx scaling as $N\mu$ is absent, significantly dampening the continual, stochastic reintroduction of upstream reciprocators.

Furthermore, because internal gratitude states are systematically reset at the beginning of every generation, upstream reciprocators remain highly vulnerable to exploitation by defectors. This emotional clearing suppresses the relative payoffs of upstream reciprocators, preventing their proliferation across large population sizes. The exception to this trend occurs at very small values of $N$, where the stationary frequency of upstream reciprocators is slightly higher. In these highly confined populations, upstream reciprocators effectively circulate cooperative actions, which bolster their average payoffs. This observation is entirely consistent with the established literature regarding the small-group effect on the persistence of upstream reciprocity~\cite{Boyd1988,Boyd1989}.

Next, we fix the population size $N = 150$ and vary the network degree $k$. As clearly shown in Fig.~\ref{fig:overlapping_population_all_update}(a)(v), we observe that the maximum frequency of upstream reciprocators lies at $k_{\max}=2$, after which it declines sharply and remains nearly zero. Concurrently, the global cooperation level rapidly collapses, and remains, entirely negligible across all higher values of $k$, while the defector fraction quickly approaches and saturates near 100\%. The presence of $k_{\max}$ at $2$ can be understood intuitively through the following microscopic arguments; for a quantified assessment, we refer the reader to Appendix~\ref{sec:kmax_network}.

As established in our discussion of the NG framework (baseline scenario), the evolutionary viability of upstream reciprocators is governed by a trade-off between two competing factors: (i) the local clustering of upstream individuals, which insulates them and sustains cooperation, and (ii) boundary exposure to defectors, which penalizes them. Under the OGG update rule, where only a small fraction of individuals are replaced in each generation, spatial structures evolve far more slowly. Consequently, cooperative clusters persist over much longer timescales compared to the volatile generation-wide sweeps of the NG framework. Individuals embedded within these long-lived clusters can repeatedly interact with the same cooperative partners over multiple generations, consistently renewing their gratitude states and accumulating higher payoffs.

Because of this heightened temporal stability, the highly isolated linear clusters that form at $k=2$ become exceptionally resilient. At this minimal degree, an upstream cluster has only two boundary points exposed to external exploitation, minimizing its losses. Consequently, the two defectors at the boundaries---whose impact is devastating in the NG case---due to generational persistence do not inflict severe damage under this update scheme. As a result, upstream reciprocators attain their peak frequency at $k=2$, yielding $k_{\max}=2$. However, as the network degree increases from $2$ to $3$ and beyond, these linear clusters dissolve into a more interconnected topology, exposing the cooperative cores to defector penetration. Compounding this exposure, the resetting of internal gratitude states at the start of each generation leaves the increasingly well-connected upstream actors highly vulnerable to exploitation. Together, these structural and gratitude updating factors grant a selective advantage to defectors, triggering the monotonic collapse of upstream reciprocity observed as $k$ increases [see Fig.~\ref{fig:overlapping_population_all_update}(a)(v)].

\subsubsection*{Full system: In the presence of downstream reciprocators}
We now construct heatmaps of the steady-state frequencies of all three behavioural types in the $(N,k)$ parameter space, as shown in Fig.~\ref{fig:overlapping_population_all_update}(b). The first observation from these distributions is that the presence of downstream reciprocators significantly enhances the frequency of upstream reciprocators. In this regime, all three behavioural types—upstream reciprocators, downstream reciprocators, and defectors—largely coexist across. This outcome is consistent with our mean-field analysis in the infinite-population limit (Sec.~\ref{sec:INF_model}), where a stable interior fixed point is analytically supported. To understand the detailed patterns in this case and compare them with the prior regimes, we again analyze the results along our two cross-sectional directions: varying $N$ under a fixed network degree, and varying $k$ for a fixed population size.

We first examine the well-mixed cross-sectional slice of Figs.~\ref{fig:overlapping_population_all_update}(b)(i)--(iv) at $k=N-1$ as a function of the population size $N$. As shown in Fig.~\ref{fig:overlapping_population_all_update}(b)(v), we find that the frequency of upstream reciprocators monotonically increases with $N$ and eventually saturates at a high value, outcompeting both downstream reciprocators and defectors. Although the frequency of downstream reciprocators decreases to some extent as $N$ expands, both upstream and downstream reciprocating types persist at levels significantly higher than defectors. Consequently, the global cooperation level remains elevated throughout this well-mixed regime.

This observed frequency expansion of upstream reciprocators with $N$ can be understood through the temporal persistence of spatial groups under OGG. Because the OGG framework updates only a small fraction of the population in each generation, heterogeneous cooperative alliances between upstream and downstream reciprocators can persist and self-reinforce over long timescales without being disrupted by generation-wide sweeps.
The ultimate dominance of the upstream strategy over the downstream strategy at large values of $N$ can be directly attributed to the underlying reputation dynamics. Downstream reciprocators aggressively suppress defectors, but in doing so, they frequently refuse to help them, which causes them to accumulate a poor public reputation. In contrast, upstream reciprocators holding a high gratitude state ($E_i = {\rm H}$) blindly continue to pass help forward, thereby maintaining a favorable reputation. Once defectors are sufficiently marginalized and controlled by the downstream population, the immaculate reputations of the upstream reciprocators hand them a relative fitness advantage. This asymmetric payoff bonus allows the upstream strategy to systematically surpass the downstream strategy as $N$ climbs into the large-population limit [Fig.~\ref{fig:overlapping_population_all_update}(b)(v)].

Notably, at smaller values of $N$, downstream reciprocators heavily dominate the population. In this low-$N$ regime, the penalty incurred by downstream enforcement is highly mitigated because the total number of interaction rounds—which scales $R = N(N-1)/2$—is small, leaving defectors with fewer opportunities to tarnish downstream reciprocators image-scores. However, when $N$ is exceptionally small, the classic small-group effect~\cite{Boyd1988,Boyd1989} briefly acts to protect upstream reciprocity. This explains the slight initial dip in the upstream frequency profile before collective, reputation-mediated effects fully take over and drive their steady-state abundance upward in larger populations.

Next, we fix the population size at $N = 150$ and vary the network degree $k$ as shown in Fig.~\ref{fig:overlapping_population_all_update}(b)(vi). We observe that the frequency of upstream reciprocators initially climbs with as we increase $k$, reaching a local maximum around $k_{\max}\approx 5$, before \emph{gradually} decaying and saturating at a finite value. This saturation plateau scales upward with larger population sizes $N$, following the identical mechanisms detailed in our varying-$N$ analysis. Consequently, the global cooperation level remains substantially higher than in the baseline scenario and shows a peak at the $k_{\max}$.

As discussed previously, four topological factors govern the emergence of this optimal connectivity peak and the overall profile shape: (i) the formation of upstream clusters, (ii) the formation of supportive downstream clusters, (iii) the cross-links bridging upstream and downstream reciprocators, and (iv) the boundary edges shared between upstream reciprocators and defectors. While the first three actively drive the evolution of upstream reciprocators, the final one imposes an evolutionary penalty. The non-monotonic interplay among these competing spatial forces determines the precise location of $k_{\max}$ and the subsequent saturation profile. For a fully quantified topological treatment, we refer the reader to Appendix~\ref{sec:kmax_network}.

Microscopically, the emergence of a peak around $k_{\max}\approx 5$ can be understood using an intuition similar to the NG framework, albeit with several distinctions. In stark contrast to the NG case, the frequency of upstream reciprocators at the one-dimensional ring limit ($k=2$) is far from negligible. Because the OGG framework updates the population partially, linear chains of upstream and downstream reciprocators can persist over much longer timescales. These temporally stable cooperative groups successfully sustain mutual assistance even when flanked by defectors along the ring. As $k$ increases from 2 toward 5 and more, the expanded local branching factor multiplies the likelihood of forming dense, mixed upstream--downstream cooperative clusters, allowing both behavioural types to proliferate through mutual fitness feedback. However, as connectivity increases far beyond $k_{\max} \approx 5$, these heterogeneous structures become increasingly exposed to defector invasion due to disassortative mixing. This breakdown of spatial insulation diminishes the cooperative payoff advantage, forcing the frequency of upstream reciprocators to gradually decay before flattening into its well-mixed saturation limit. The gradual nature of this decay, as opposed to a rapid collapse, stems directly from the long-term persistence of local groups under overlapping generations.

\subsection{Overlapping generation with local resetting (OGL): Only newly born individuals' image-score and gratitude reset in every generation}\label{sec:overlapping_Only_new_Update}

\begin{figure*}
	\includegraphics[scale=1.00]{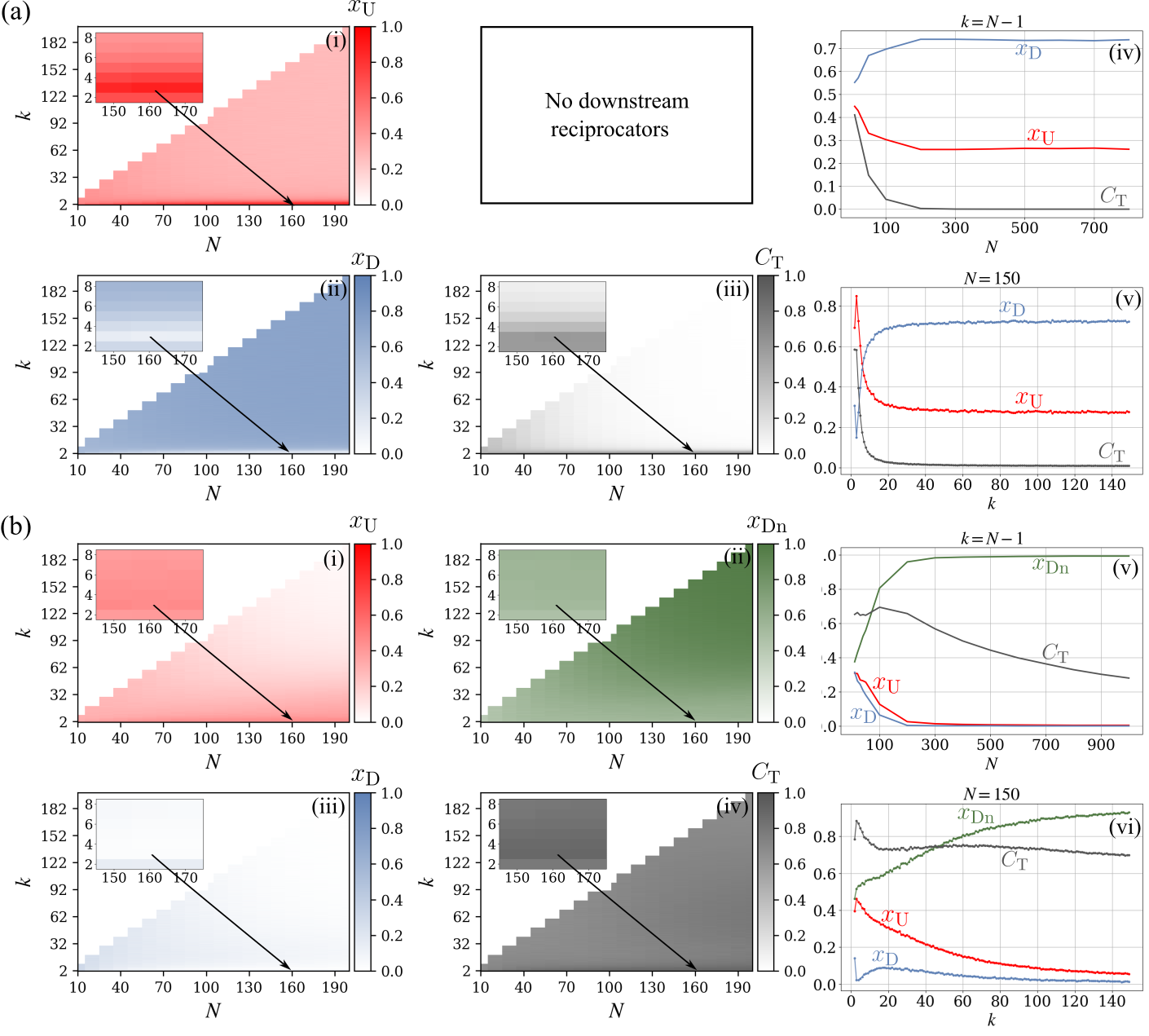}
	\caption{Steady-state frequency distributions and cooperation level under the overlapping generation with local resetting (OGL): Results are given as a function of the population size $N$ and network degree $k$. Panel~(a) depicts the outcomes in the absence of downstream reciprocators (baseline scenario), while panel~(b) exhibits the results when all three behavioural types are present (full system). In both major blocks, the heatmaps shows the long-term steady-state frequencies ($x_{\rm U}$, $x_{\rm Dn}$, $x_{\rm D}$) and cooperation level ($C_{\rm T}$) across the $(N, k)$ parameter space. The insets in (a)(i)--(iii) and (b)(i)--(iv) shows magnified views of the small-$k$ region near the $N$ value indicated by the arrow. Panels~(a)(iv) and (b)(v) show the dependence of strategy frequencies on $N$ alone, with the degree fixed at $k = N-1$. Panels~(a)(v) and (b)(vi) show the dependence on $k$ alone, with population size fixed at $N= 150$.}
	\label{fig:overlapping_pop_new_born_update}
\end{figure*}

 We compute the steady-state frequency distributions across the $(N,k)$ parameter space for OGL scenario, as illustrated in Fig.~\ref{fig:overlapping_pop_new_born_update}.

\subsubsection*{Baseline scenario: In the absence of downstream reciprocators}

As before, we begin our analysis with the baseline scenario where only upstream reciprocators and defectors are present. The steady-state frequency distributions across the full $(N,k)$ parameter space are displayed in Figs.~\ref{fig:overlapping_pop_new_born_update}(a)(i)--(iii), while the explicit cross-sectional cuts along the $N$- and $k$-directions are systematically explored in Fig.~\ref{fig:overlapping_pop_new_born_update}(a)(iv) and Fig.~\ref{fig:overlapping_pop_new_born_update}(a)(v), respectively. In contrast to all preceding frameworks, we observe a fundamentally different trend: even in the total absence of downstream reciprocators, the frequency of upstream reciprocators remains remarkably substantial [Fig.~\ref{fig:overlapping_pop_new_born_update}(a)(iv)]. In fact, among the three generational configurations (NG, OGG, and the present setup, i.e., OGL) investigated in this work, this scenario yields the highest abundance of the upstream reciprocators in the baseline scenario.

The reason behind this elevated upstream frequency [Fig.~\ref{fig:overlapping_pop_new_born_update}(a)(iv)] is quite intuitive. Under this framework, the gratitude state reset to $E_i = {\rm H}$ is strictly localized to newly born individuals. Consequently, the vast majority of upstream reciprocators that transition into the ungrateful state ($E_i = {\rm L}$) via initial contact with defectors remain trapped in that state for extended periods. This creates an evolutionary regime where a massive fraction of the upstream population effectively functions as defectors. As a result, the payoff difference between these two behavioural types virtually vanishes, allowing the upstream strategy to persist under this weakened selection pressure. This interpretation is directly validated by the cooperation level profile, which remains heavily suppressed near zero across large population sizes. It is only with small $N$ that upstream reciprocators can effectively propagate aid within small local groups before being overwhelmed by defectors, momentarily lifting the cooperation level. This striking observation demonstrates that while the upstream behavioural type can survive within a defector population, its evolutionary viability and behavioural expression heavily depends upon the microscopic update rule.

When we isolate the cross-sectional slices [of Figs.~\ref{fig:overlapping_pop_new_born_update}(a)(i)--(iii)] across the network degree $k$ while keeping the population size fixed at $N = 150$, we observe that the frequency of upstream reciprocators reaches a distinct local maximum at $k_{\max}=3$ as illustrated in Fig.~\ref{fig:overlapping_pop_new_born_update}(a)(v). At this maximum, the upstream population dramatically peaks and even surpasses the abundance of defectors. Beyond this maximum, the frequency of upstream reciprocators gradually decreases and eventually flattens into a finite saturation plateau. Interestingly, the global cooperation level follows a distinct trend: Even though it also reaches a maximum near $k=3$, but subsequently plummets to nearly zero in denser networks.

The origin of this non-monotonic pattern can once again be mapped to the interplay between two competing quantities: (i) the number of upstream clusters and (ii) the number of edges shared with defectors (see Appendix~\ref{sec:kmax_network} for a quantified analysis). Within this OGL framework, upstream clusters persist for longer times because population are no longer globally wiped out at each generational transition. Consequently, individuals embedded deep within these clusters can repeatedly cooperate over extended timescales, sustaining localized mutual assistance and accumulating higher payoffs. Moreover, upstream reciprocators situated along the vulnerable boundaries of these clusters transition into the ungrateful state ($E_i = {\rm L}$) upon their initial interaction with external defectors. Because their emotional states do not reset in each generation, these boundary-dwelling, ungrateful upstream actors become immune to further exploitation, effectively acting as a protective shield around the cooperative cluster.

As $k$ increases past the maximum, these clusters are subjected to mixing, increasing their exposure to defectors and ultimately eroding the cooperative clusters. However, the delicate balance between clustering and mixing shifts the maximum to an intermediate connectivity of $k_{\max}=3$ rather than the absolute minimal degree of $k=2$. This occurs because the higher coordination number allows for the formation of larger, upstream clusters that can be insulated by a protective perimeter of ungrateful, exploitation-resistant boundary actors, optimizing the cluster's collective survival before dissolution takes over in well-connected networks.

\subsubsection*{Full system: In the presence of downstream reciprocators}

In the full system, where all three behavioural types are present, we observe a fundamentally different qualitative pattern compared to the previous two generational architectures (NG and OGG), as evident from the steady-state frequency distributions shown in Fig.~\ref{fig:overlapping_pop_new_born_update}(b). Unlike the preceding frameworks, the first striking feature of these heatmaps is that downstream reciprocators no longer act as an evolutionary catalyst that promotes the persistence of upstream reciprocators; instead, they exert a diminishing effect on the upstream frequency. Indeed, the downstream reciprocators outcompetes the other two behavioural types, dominating the vast majority of the $(N,k)$ parameter space, as shown in Figs.~\ref{fig:overlapping_pop_new_born_update}(b)(i)--(iv). To systematically unpack the microscopic mechanisms driving this change, we explore the dependencies along our two cross-sectional axes ($N$ and $k$) separately.

We first analyze the dependence of the three behavioural types on the population size $N$ along the well-mixed cross-section ($k=N-1$) of Fig.~\ref{fig:overlapping_pop_new_born_update}(i)--(iv). As shown in Fig.~\ref{fig:overlapping_pop_new_born_update}(b)(v), upstream reciprocators and defectors are present in appreciable proportions only for small values of $N$. As $N$ increases, however, the stationary frequencies of both upstream reciprocators and defectors decline monotonically, while downstream reciprocators expand to dominate the entire population. This outcome is in full alignment with the infinite-population analysis presented in Sec.~\ref{sec:INF_model}, where---depending on the parameter space, specifically within regions characterized by a low fraction of good reputation upstream reciprocators ($g_{\mathrm U}$) and a high fraction of reputable downstream reciprocators ($g_{\mathrm{Dn}}$)---the fixed point corresponding to downstream dominance is stable.

The reason behind this pattern in Fig.~\ref{fig:overlapping_pop_new_born_update}(b)(v) lies in the absence of a global generation-wide update of public reputations and internal gratitudes. Because the reset to a favorable public reputation and a high gratitude state ($E_i = {\rm H}$) is strictly localized to newly born individuals, adult upstream reciprocators that lapse into the ungrateful state ($E_i = {\rm L}$) due to interactions with defectors, and hence defect subsequently, acquire a bad public reputation that persists over long periods. Consequently, downstream reciprocators---unable to distinguish between an actual defector and an ungrateful upstream reciprocators based on reputation alone---withholding cooperative benefits from both. This collateral damage strips away the relative payoff advantage of the upstream reciprocators, severely suppressing their frequency.

This mechanism is the opposite of the OGG framework, where periodic, global resetting of public reputations routinely erased such long-term reputational penalties. Under the informational inheritance scheme in the current OGL case, since only a small fraction of individuals are replaced in each generation, historical records linger in the adult population, disadvantaging both defectors and ungrateful upstream reciprocators alike. As $N$ increases, the total interactions expands ($R=N(N-1)/2$), amplifying this reputational filtering and driving the frequencies of both upstream reciprocators and defectors toward extinction. This also explains the concurrent decline in the cooperation level with increasing $N$, as cooperative acts become progressively rarer. Overall, these results highlight the vital importance of timely reputational updates and emotional refreshing for sustaining large-scale social cooperation.

Finally, we analyze the dependence of the frequencies on the network degree $k$ while keeping the population size fixed at $N = 150$. As shown in Fig.~\ref{fig:overlapping_pop_new_born_update}(b)(vi), we observe the emergence of a peak at $k_{\max}=3$, where the frequency of upstream reciprocators is maximized. Beyond this optimal degree, the abundance of upstream reciprocators decays steadily, whereas the frequency of downstream reciprocators expands. In highly connected networks, strong disassortative mixing allows downstream reciprocators to outcompete the other behavioural types and dominate the population.

As discussed in the previous cases, the emergence of the peak at $k_{\max}$ is governed by the interplay of the four competing factors: (i) upstream clusters, (ii) downstream clusters, (iii) cross-links bridging upstream and downstream reciprocators, and (iv) edges shared with defectors (see Appendix~\ref{sec:kmax_network} for a quantified analysis). However, unlike the preceding generational architectures, interaction links between upstream and downstream reciprocators may now exert a diminishing effect on the payoff of the upstream reciprocators. Because public reputations and internal gratitude states are no longer globally updated at each generational transition, adult upstream reciprocators that carry a poor public reputation persists over long periods. Consequently, downstream reciprocators withhold help from these ungrateful upstream actors. Far from acting as a promoter as it did in the previous updating regimes, factor (ii) and (iii) no longer guarantees a positive fitness feedback loop for the upstream population. Remarkably, our analysis in Appendix~\ref{sec:kmax_network} confirm that when the interaction links between upstream and downstream reciprocators along with downstream clusters are omitted, the calculated peak matches our observed value of $k_{\max}=3$ exactly.

\section{Comprehending the results}\label{sec:comphrendhing}
In this section, we collectively compare the results obtained under the three generational update rules investigated in the previous section. By systematically examining these frameworks side by side, we aim to gain deeper insight into the observed phenomena and identify the mechanisms underlying the distinct evolutionary outcomes. 

A collective comparison of our three generational frameworks---namely, non-overlapping generations (NG), overlapping generations with global resetting (OGG), and overlapping generations with local resetting (OGL)---yields foundational insights into the mechanisms governing the evolution of upstream reciprocity. The most robust observation across all cases is that the introduction of downstream reciprocators enhances the cooperation level relative to the baseline scenario (where only upstream reciprocators and defectors are present). This highlights the universality of downstream reciprocity in catalyzing helping behavior across fundamentally different generational architectures.

However, the effect of downstream reciprocators on the persistence of upstream reciprocators depends strongly on the underlying update rule. In the NG and OGG cases, the introduction of downstream reciprocators acts as an evolutionary promoter, substantially enhances the frequency of upstream reciprocators. In contrast, under OGL, downstream reciprocators have a diminishing effect on upstream population. This difference arises because, in OGL, the absence of global updating of image-scores and gratitude causes low-gratitude upstream reciprocators to accumulate persistent bad reputations. As a result, downstream reciprocators indiscriminately withhold help from both, inflicting collateral reputational damage that heavily suppresses upstream growth. 

Interestingly, hints of these contrasting outcomes are already present in the infinite-population model (Sec.~\ref{sec:INF_model}). One possible outcome is the coexistence of all three behavioural types, which is consistent with the patterns observed under NG and OGG. Another possible outcome is the dominance of downstream reciprocators, which closely resembles the behavior observed in the OGL case. This suggests that the different update rules in structured populations selectively favor different equilibrium tendencies already embedded in the system's underlying mean-field dynamics.

This comparison also highlights the evolutionary significance of gratitude updating. In the absence of downstream reciprocators, upstream reciprocators are present in appreciable proportions only in OGL. However, the cooperation level in this case remains paradoxically low, indicating that a large fraction of upstream reciprocators are trapped in the ungrateful state and therefore behave effectively as defectors. This demonstrates that a high frequency of upstream reciprocators alone does not automatically translate to widespread cooperative behavior. Rather, one needs regular gratitude resetting for upstream reciprocators to propagate cooperation.

Another striking observation common to all cases is the existence of an optimal degree, $k_{\max}$, at which the frequency of upstream reciprocators along with cooperation level reaches a maximum. We show that this optimum emerges from a tug of war between cooperative clustering and mixing with defectors. The coexistence of high upstream frequency and high cooperation near $k_{\max}$ suggests that a large fraction of upstream reciprocators are in a high-gratitude state. This highlights the crucial role that network structure can play: By limiting the number of neighboring individuals with whom one interacts, network structure facilitates the persistence of high-gratitude upstream reciprocators.

Finally, the dependence on population size $N$ differs significantly across update rules. In the absence of downstream reciprocators, the frequency of upstream reciprocators increases with $N$ only in the NG case. When all three behavioural types are present, however, upstream reciprocators increase with $N$ in both NG and OGG, but not in OGL. These contrasting trends further emphasize the importance of the update mechanism, particularly the role of reputation and gratitude resetting, in determining the persistence of upstream reciprocity and cooperative behavior in large populations.

\section{Conclusion}\label{sec:conclusion}
Two individuals help a stranger. One does so because of a kindness once received; the other, in quiet anticipation of one yet to come. Though their intrinsic motivations differ, the question of whether either survives the pressure of natural selection is anything but simple. Indeed, survival in an evolving population is shaped not only by individual intent, but by the invisible architecture surrounding it---the structure of the network, the dynamics of experiential resetting, and the currency of reputation.

A central finding of our work is that the evolutionary fate of upstream reciprocity depends sensitively on the underlying update mechanism, a subtle dimension often overlooked in the literature. We demonstrate that global versus local updating of reputation and gratitude can fundamentally alter evolutionary outcomes, ranging from coexistence between behavioural types to the dominance of reputation-based downstream reciprocators. At the same time, we uncover a robust structural feature across all update rules: the emergence of an optimal network degree, $k_{\max}$, at which upstream reciprocity is maximized. This optimum arises from a balance between cooperative clustering and exposure to defectors, highlighting the nontrivial role of network structure in sustaining cooperation.

Taken together, these results paint a picture that is richer---and more nuanced---than either mechanism alone would suggest. More broadly, our findings suggest that comprehensively decoding the puzzle of social cooperation requires not only identifying who interacts with whom, but also how social memory, reputation, and internal states evolve over time. In this sense, the present work is perhaps less an endpoint than a beginning toward a more complete understanding of how, the joint evolution of multiple forms of reciprocity in realistic social systems take place.

With that spirit in mind, we briefly sketch some directions that this beginning points toward. An immediate question concerns the role of private information~\cite{Hilbe2018,Nowak1998a}: in our model, all the reputations are publicly available, whereas real social systems are often characterized by private and possibly conflicting assessments of the same individual. How such informational heterogeneity alters the delicate evolutionary balance between upstream and downstream reciprocity remains an open question. Closely related is the role of higher-order social norms, where even the act of punishing an individual may itself be judged differently~\cite{Ohtsuki2004}.

In a similar spirit, relaxing the assumption of binary memory---allowing individuals to track not merely whether someone is good or bad, but the exact degree to which they are good---may reveal a richer and more realistic landscape of cooperation~\cite{Nowak1998a}. Beyond information structure, the network itself deserves further scrutiny: adaptive and more complex networks, where links form and disappear in response to behavior, may fundamentally reshape the persistence of reciprocity~\cite{Zimmermann2004,Pacheco2006}.
Extending this framework to multiplayer interaction settings~\cite{Broom1997,Dubey2026_chaos}, such as public goods games~\cite{Patra_2022,Sasaki2026}, would further connect these ideas to the collective action problems that characterize much of humans and animals social life.

\section*{Data Availability}

The data and code supporting the findings of this study are available at GitHub: \url{https://github.com/vdubey9818/upstream_vs_downstream/tree/main}.

\appendix
\section{Quantitative explanation of the emergence and location of $k_{\max}$}\label{sec:kmax_network}

In this section, we quantitatively examine the mechanisms underlying the emergence of $k_{\max}$ across all three generational configurations (NG, OGG and OGL). As discussed in the main text, this peak arises from a fundamental trade-off between the two competing factors in the baseline case (with U and D only) and four factors in the full system (with U, Dn and D). All these factors can be broadly grouped into two classes.  The first is the local clustering of upstream reciprocators, downstream reciprocators, and defectors, which facilitates the formation of intra-type communities that effectively function as small, partially isolated sub-populations. The second is the set of boundary edges interconnecting different behavioural types, which dictates the rate of disassortative mixing between them. Here, we formalize and quantify these competing effects and show that the maxima of our computed structural metrics precisely align with the empirical values of $k_{\max}$ observed in the main text.

To quantify this first class of factors, namely the characteristic size of these homogeneous behavioural clusters, we implement the following graph-theoretic procedure. At each generation $t$, we employ the Breadth-First Search (BFS) algorithm~\cite{Barabasi2016,Moore1959} to determine two properties of the network at each generation: (i) the total number of disconnected clusters composed entirely of a single behavioural type, denoted by $z_{\rm U}(k,t)$, $z_{\rm Dn}(k,t)$, and $z_{\rm D}(k,t)$ for upstream reciprocators, downstream reciprocators, and defectors, respectively; and (ii) the sizes of these clusters, where the size of the $i$-th cluster is denoted by $s_{\rm U}^i(k,t)$, $s_{\rm Dn}^i(k,t)$, and $s_{\rm D}^i(k,t)$. Using these quantities, we calculate the average cluster size of each behavioural type by taking a time average over $T$ generations:
\begin{subequations}\label{eqn:cluster}
	\begin{eqnarray}
		S_{\rm U}(k) &=& \frac{1}{T}\sum_{t=1}^{T}\frac{1}{z_{\rm U}(k,t)}\sum_{i=1}^{z_{\rm U}(k,t)} s_{\rm U}^i(k,t),\\
		S_{\rm Dn}(k) &=& \frac{1}{T}\sum_{t=1}^{T}\frac{1}{z_{\rm Dn}(k,t)}\sum_{i=1}^{z_{\rm Dn}(k,t)} s_{\rm Dn}^i(k,t),\\
		S_{\rm D}(k) &=& \frac{1}{T}\sum_{t=1}^{T}\frac{1}{z_{\rm D}(k,t)}\sum_{i=1}^{z_{\rm D}(k,t)} s_{\rm D}^i(k,t).
	\end{eqnarray}
\end{subequations}
These quantities represent the time-averaged cluster size of each behavioural type in a random regular network with degree $k$.

To quantify the second one, namely the extent of disassortative mixing between behavioural types, we numerically calculate the number of edges connecting different phenotypes in each generation $t$. Specifically, we measure the number of edges between upstream reciprocators and defectors, $e_{\rm U\text{-}D}(k,t)$; upstream and downstream reciprocators, $e_{\rm U\text{-}Dn}(k,t)$; and downstream reciprocators and defectors, $e_{\rm Dn\text{-}D}(k,t)$. Using these quantities, we compute the average number of edges by taking a time average over $T$ generations:
\begin{subequations}\label{eqn:egdes}
	\begin{eqnarray}
		E_{\rm U\text{-}D}(k) &=& \frac{1}{T}\sum_{t=1}^{T} e_{\rm U\text{-}D}(k,t),\\
		E_{\rm U\text{-}Dn}(k) &=& \frac{1}{T}\sum_{t=1}^{T} e_{\rm U\text{-}Dn}(k,t),\\
		E_{\rm Dn\text{-}D}(k) &=& \frac{1}{T}\sum_{t=1}^{T} e_{\rm Dn\text{-}D}(k,t).
	\end{eqnarray}
\end{subequations}

\begin{figure*}
	\includegraphics[scale=0.80]{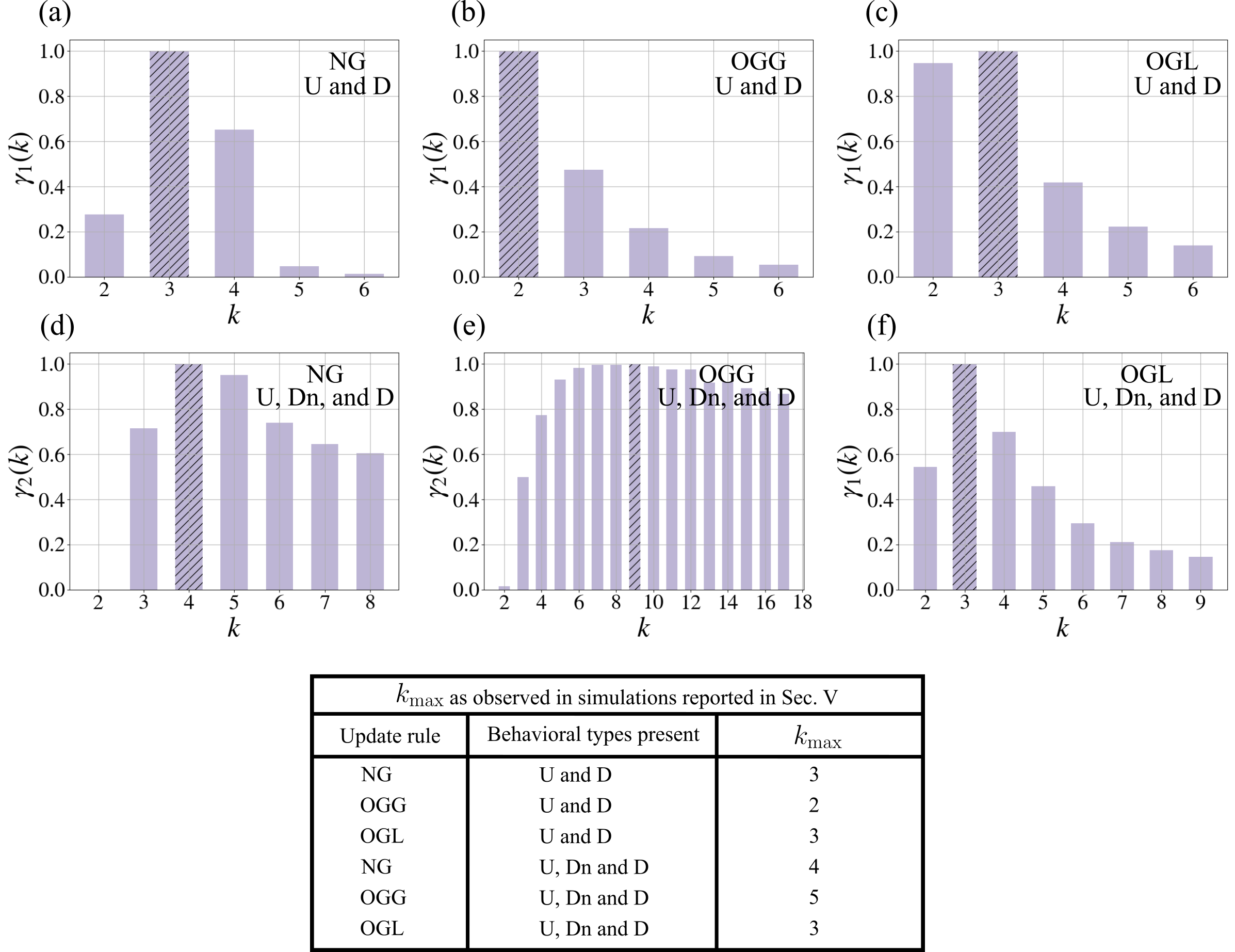}
	\caption{Effective clustering measures as functions of degree, illustrating that the maxima of these measures occur at the corresponding values of $k_{\max}$ observed in simulation reported in Sec.~\ref{sec:Results}. For reference we have tabulated all the $k_{\max}$'s in the table (below the figures) for each of the scenarios. In each panel the hatched histogram indicate $\max_k\{\gamma_i(k)\}$ for $i\in\{1,2\}$. Subplots (a), (b), and (c) show the effective measure $\gamma_1(k)$ for the baseline scenario under the three update rules NG, OGG and OGL respectively. Subplots (d) and (e) illustrate the measure $\gamma_2(k)$ for NG and OGG cases for the three-type full system. Finally, subplot (f) depicts the measure $\gamma_1(k)$ for the three-type full system under the OGL update rule.}
	\label{fig:clustering}
\end{figure*}

As discussed in the main text, these two factors can have opposing effects. For instance, in the baseline scenario where only upstream reciprocators and defectors are present, the clustering of upstream reciprocators [$S_{\rm U}(k)$], promotes the persistence of the same, whereas the edges between upstream reciprocators and defectors [$E_{\rm U\text{-}D}(k)$], suppress it.
Motivated by this competition, we define the following effective clustering measure to mathematically characterize the non-monotonic behavior of the upstream population:
\begin{equation}\label{eq:gamma1}
	\Gamma_1(k)=\frac{S_{\rm U}(k)}{E_{\rm U\text{-}D}(k)}.
\end{equation}
For visual clarity and seamless comparison across our different generational architectures, we utilize the normalized form of this structural ratio:
\begin{equation}\label{eq:gamma1_norm}
	\gamma_1(k)=\frac{\Gamma_1(k)}{\max_k\{\Gamma_1(k)\}}.
\end{equation}
As shown in Figs.~\ref{fig:clustering}(a)--(c),  we plot the normalized effective clustering coefficient $\gamma_1(k)$ as a function of the network degree $k$ across our three update frameworks (NG, OGG, and OGL). Remarkably, we observe that the maxima of $\gamma_1(k)$ occur exactly at the corresponding values of $k_{\max}$ for each case as evident from the table in Fig.~\ref{fig:clustering}.

The case where all three behavioural types are present is somewhat more subtle, as downstream reciprocators play an important role in shaping the emergence of upstream reciprocators. As discussed in the main text, downstream clusters enhance the persistence of upstream reciprocators. Similarly, edges between upstream and downstream reciprocators, quantified by $E_{\rm U\text{-}Dn}(k)$, also have a positive effect, whereas edges between upstream reciprocators and defectors suppress upstream persistence. Motivated by these competing effects, we define the following effective clustering measure:
\begin{equation}
	\Gamma_2(k)=\frac{S_{\rm U}(k)\, S_{\rm Dn}(k)\, E_{\rm U\text{-}Dn}(k)}
	{E_{\rm U\text{-}D}(k)}.
\end{equation}
As before, for visual clarity and seamless comparison across the different setups, we employ the normalized form of this clustering measure:
\begin{equation}
	\gamma_2(k)=\frac{\Gamma_2(k)}{\max_k\{\Gamma_2(k)\}}.
\end{equation}
As shown in Fig.~\ref{fig:clustering}(d), we plot $\gamma_2(k)$ as a function of $k$ for the NG case in the populations where all three behavioural types are present. We observe that the maxima of $\gamma_2(k)$ coincide exactly with the corresponding values of $k_{\max}$. However, the case of OGG and OGL frameworks are different and are examined individually below.

In the OGG case, $k_{\max}$ lies around 5, whereas here the maximum of $\gamma_2(k)$ lies around 9 [as shown in Fig.~\ref{fig:clustering}(e)], indicating a slight quantitative mismatch. Nevertheless, as evident from comparing Fig.~\ref{fig:overlapping_population_all_update}(b)(vi) and Fig.~\ref{fig:clustering}(e), the qualitative trend in the vicinity of these peaks remains identical: both profiles exhibit a matching, gradual variation around their respective maxima. Notably, the deviation between $\gamma_2(k_{\max})$ and $\max_{k}\{\gamma_2(k)\}$ is remarkably small. We attribute this minor discrepancy to stochastic fluctuations and higher-order structural corrections—such as those arising from Dn–Dn and Dn–D interactions—which are not explicitly incorporated into our clustering measure. These omitted interactions introduce corrections that slightly perturb the underlying trend without fundamentally altering it.

The OGL case is significantly different compared to the previous two cases (NG and OGG), as here,upstream reciprocators face a penalty from defectors that is compounded by a more ambiguous interaction with downstream reciprocators. As discussed in the main text, downstream reciprocators may punish low-gratitude upstream reciprocators whose bad image persists over time due to the absence of global gratitude resetting. 
However, unlike the unconditional penalty imposed by defectors, this effect is not guaranteed: At low degree, upstream reciprocators can have either high or low gratitude, depending on whether they reside within upstream clusters or at the boundary with defectors. Consequently, whether downstream reciprocators harm or benefit upstream reciprocators depend on the position on the network. Given this uncertainty, we retain the contribution of edges $E_{\rm U\text{-}Dn}(k)$ and downstream groups $S_{\rm Dn}(k)$ in the effective clustering metric. This allows $\gamma_1(k)$, as defined in Eq.~(\ref{eq:gamma1}) and Eq.~(\ref{eq:gamma1_norm}), to reflect the situation in this case.

As shown in Fig.~\ref{fig:clustering}(f), the maximum of $\gamma_1(k)$ as a function of $k$ coincides exactly with the observed value of $k_{\max}$ in the table in Fig.~\ref{fig:clustering}. Moreover, we observe that the qualitative pattern near $\max_{k}\{\gamma_1(k)\}$ closely matches the trend near $k_{\max}$ in Fig.~\ref{fig:overlapping_pop_new_born_update}(b)(vi).
This flawless alignment successfully completes our quantitative assessment, providing a robust mathematical foundation that reinforces the macroscopic intuition detailed in the main text.

\bibliography{Dubey_etal_bibliography.bib}	
\end{document}